\documentclass[11pt, a4paper]{article}

\usepackage{graphicx}
\usepackage{bm}
\usepackage{amsmath, amssymb,theorem,verbatim}
\usepackage[round]{natbib} 
\usepackage{setspace}
\usepackage{multirow} 
\usepackage{hyperref}
\usepackage{bm}
\usepackage{algpseudocode}
\usepackage{textcomp}

\long\def\comment#1{}

\newtheorem{theorem}{Theorem}[section]

\newtheorem{prop}{Proposition}[section]

\newtheorem{cor}{Corollary}[section] 

\newtheorem{defin}{Definition}[section]

\newtheorem{assumption}{Assumption}[section]

\begin{document}

\doublespacing

\author{Piotr Fryzlewicz\thanks{Department of Statistics, London School of Economics, Houghton Street, London WC2A 2AE, UK. Email: \href{mailto:p.fryzlewicz@lse.ac.uk}{p.fryzlewicz@lse.ac.uk}.}}

\title{Robust Narrowest Significance Pursuit: Inference for multiple change-points in the median}





\oddsidemargin=0.25in
\evensidemargin=0in
\textwidth=6in
\headheight=0pt
\headsep=0pt
\topmargin=0in
\textheight=9in

\maketitle

\begin{abstract}
We propose Robust Narrowest Significance Pursuit (RNSP), a methodology for detecting localized regions in data sequences which each must contain a change-point in the median, at a prescribed global significance level.
RNSP works by fitting the postulated constant model over many regions of the data using a new sign-multiresolution sup-norm-type loss, and greedily
identifying the shortest intervals on which the constancy is significantly violated. By working with the signs of the data around fitted model candidates, RNSP fulfils its coverage promises under minimal assumptions, requiring only sign-symmetry and serial independence of the signs of the true residuals. In particular, it permits their heterogeneity and arbitrarily heavy tails. The intervals of significance returned by RNSP have a finite-sample character, are unconditional in nature and do not rely on any assumptions on the true signal. Code
implementing RNSP is available at \url{https://github.com/pfryz/nsp}.

\vspace{5pt}

\end{abstract}

\section{Introduction}

The problem of uncertainty quantification for possibly multiple parameter changes in time- or space-ordered data is motivated
by the practical question of whether any suspected changes reflect real structural changes in the underlying stochastic model, or are
due to random fluctuations. Approaches to this problem include confidence sets associated with simultaneous multiscale change-point estimation
\citep{fms14, psm17, vbm22}, post-selection inference \citep{hgst18, hlgst18, jfw20, dtst20}, inference without selection and post-inference selection via Narrowest Significance Pursuit \citep{f21},
asymptotic confidence intervals conditional on the estimated change-point locations \citep{bp98, km11, ck21}, False Discovery Rate \citep{hnz13, lm16, chs19}, and Bayesian inference \citep{ll01, f06}.
These approaches go beyond mere change-point detection and offer statistical significance statements regarding the existence and locations of change-points in the statistical model underlying the data.


In this paper, we are concerned with the following problem: given a sequence of noisy observations, automatically determine localized regions of the data which each must contain a change-point in the median, at a prescribed global significance level $\alpha$. The methodology we introduce, referred to as Robust Narrowest Significance Pursuit (RNSP), achieves this for the piecewise-constant median model, capturing changes in the level of the median.
By its algorithmic construction, RNSP offers exact finite-sample coverage guarantees with practically no distributional assumptions on the data, other than serial independence of the signs of the true residuals
and their sign-symmetry, a weak requirement which is immaterial for continuous distributions (as all, even non-symmetric, continuous distributions are sign-symmetric). In contrast to the existing literature, RNSP requires no knowledge of the distribution on the part of the user, and permits arbitrarily heavy tails, heterogeneity over time and/or lack of symmetry.
In addition, RNSP is able to handle distributions that are continuous, continuous with mass points, or discrete, where these properties may also vary over the signal.
The execution of RNSP does not rely on having an estimate of the number or locations of change-points. Critical values needed by RNSP do not depend on the noise distribution, and can be accurately approximated analytically. RNSP explicitly targets the shortest possible intervals of significance. It is worth noting, however, that our large-sample consistency result for RNSP, shown in Section \ref{sec:lsb}, relies on stronger assumptions than its finite-sample coverage properties, discussed in Sections \ref{sec:mot} and \ref{sec:rnsp}. We now
situate RNSP in the context of the existing literature.

Heterogeneous Simultaneous Multiscale Change Point Estimator, abbreviated as H-SMUCE \citep{psm17}, an extension of SMUCE \citep{fms14}, is a change-point detector in the heterogeneous Gaussian piecewise-constant model $Y_i = f_i + \sigma_i \varepsilon_i$, where $f_i$ is a piecewise-constant signal, $\varepsilon_i$ are i.i.d. $N(0,1)$ variables, and $\sigma_i$ can only change when $f_i$ does.
In Section A.1 of their work, the authors provide an algorithmic construction of confidence intervals for the locations of the change-points in $f_i$, which involves
 screening the data for short intervals over which a constant signal fit is unsuitable and they must therefore contain change-points.
Crucially, this algorithmic construction relies on the knowledge of scale-dependent critical values (for measuring the unsuitability of a locally constant fit), which are not available analytically but only by simulation, and therefore the method does not extend automatically to unknown noise distributions (as the analyst needs to know what distribution to sample from). In Section \ref{sec:num}, we show that H-SMUCE suffers from inflated type I error rates in the sense that the thus-constructed confidence intervals, in the examples of Gaussian models shown, do not all contain at least one true change-point each in more than $100(1-\alpha)\%$ of the cases, contrary to what this algorithmic construction sets out to do. H-SMUCE is also prone to failing if the model is mis-specified, e.g. if the distribution of the data has a mass point (which is unsurprising in view of its assumption of Gaussianity).

Multiscale Quantile Segmentation \citep[MQS]{vbm22}, another extension of SMUCE, is a procedure for detecting possibly multiple changes in a given piecewise-constant quantile of the input data sequence, which includes the median as a special case.
MQS
estimates the number of change-points in the quantile function as the minimum among those candidate fits for which the empirical residuals pass a certain multiscale test at level $\alpha$, where the empirical residuals are defined as binary exceedance sequences of the data over the level defined by each candidate fit. Working with such binary exceedance sequences means that MQS makes no distributional assumptions on the data other than their serial independence. Like SMUCE and H-SMUCE, MQS then defines a confidence set around the estimated signals as a set of all feasible (in the same sense) signal fits at level $\alpha$. This then enables the conceptual and algorithmic construction of asymptotic simultaneous confidence intervals for the change-point locations, which are guaranteed to (each) contain a change-point with probability $1 - \alpha + o(1)$. \cite{css14} provide a critique of SMUCE from the inferential point of view, which also applies to H-SMUCE and MQS. In Section \ref{sec:num}, we illustrate the advantages and disadvantages of MQS, as implemented in the R package \verb+mqs+.
In contrast to MQS, the coverage guarantees offered by RNSP are wholly based on exact inequalities and therefore hold for any sample size, which is reflected in its performance shown in Section \ref{sec:num}.

The (non-robust) Narrowest Significance Pursuit (NSP; \citeauthor{f21}, \citeyear{f21}) is able to handle heterogeneous data in a variety of models, including the piecewise-constant signal plus independent noise model. However, NSP requires that the noise, if heterogeneous, is within the domain of attraction of the normal distribution and is symmetric, neither of which is assumed in RNSP.
The self-normalised statistic used in NSP includes a term resembling an estimate of the local variance of the noise which, however, is only unbiased under the null hypothesis of no change-point being present locally. The fact that the same term over-estimates the variance under the alternative hypothesis, reduces the power of the detection statistic, which leads to typically long intervals of significance. We illustrate this issue 
in Section \ref{sec:num} and show that RNSP offers a significant improvement.

\comment{The Bayesian approach of \cite{f06} provides an inference framework for independent observations drawn from a $f(\cdot | \theta_j)$ density within the $j$th segment, with $\theta_j$ possibly vector-valued, which permits some types of heterogeneity.
The distribution family $f$ is assumed known, and a prior need to be specified for the $\theta_j$ parameters. While self-contained and not overly computationally intensive, this approach suffers from the fact that the estimation of the number of change-points is a difficult statistical problem and therefore so is choosing a good prior for this quantity; for the same reason, it is a choice that can significantly affect the inferential conclusions.}

\cite{bp98, bp03}, working with least-squares estimation of multiple change-points in regression models under possible heterogeneity of the errors, describe a procedure for computing confidence intervals
conditional on detection, with asymptotic validity guarantees.
\comment{For an unknown distribution of the errors, the limiting distribution of each estimated change-point location converges to an appropriate functional of the Wiener process only under the assumption that the corresponding break size goes to zero with the sample size. The asymptotic validity of the resulting confidence interval relies on the consistency of the estimators of the unknown quantities involved (such as the local variance of the innovations or the break size); it is therefore a large-sample, asymptotic construction.}
Crucially, it does not take into the account the uncertainty associated with detection, which can be considerable especially for the more difficult problems (for an example, see the ``US ex-post real interest rate" case study in \cite{bp03}, where there is genuine uncertainty between models with 2 and 3 change-points; we revisit this example in Section \ref{sec:expost}). By contrast, RNSP produces intervals of significant change in the median that are not conditional on detection and have a finite-sample nature.
\comment{ and are valid regardless of the size of the breaks.}

The paper is organised as follows. Section \ref{sec:mot} motivates RNSP and sets out its general algorithmic framework.
Section \ref{sec:rnsp} describes how RNSP measures the local deviation from model constancy and gives finite-sample theoretical performance guarantees for RNSP. 
Section \ref{sec:lsb} quantifies the large-sample behaviour of RNSP.
Section \ref{sec:num} contains numerical examples and comparisons. Section \ref{sec:data} includes examples showing the practical usefulness of RNSP. Section \ref{sec:disc} concludes with a brief discussion. 
 Software implementing RNSP is available at \url{https://github.com/pfryz/nsp}.
\section{Motivation and review of RNSP algorithmic setting}
\label{sec:mot}

\subsection{RNSP: context and modus operandi}
\label{sec:mo}

RNSP discovers regions in the data in which the {\em median}
departs from constancy, at a certain global significance level. This is in contrast to NSP \citep{f21}, which targets the {\em mean}. RNSP does
not make moment assumptions about the data, and therefore the median is a natural
measure of data centrality.
We now review the components of the algorithmic framework that are shared between NSP and RNSP, 
with the generic measurement of local deviation from the constant model as one of its building blocks. In Section \ref{sec:rnsp}, we introduce the 
particular way in which
local deviation from the constant model is measured in RNSP, which is
appropriate for the median and hence fundamentally different from NSP.

RNSP operates in the signal plus noise model
\begin{equation}
\label{eq:signoise}
Y_t = f_t + Z_t, \quad t = 1, \ldots, T,
\end{equation}
in which the signal $\{  f_t   \}_{t=1}^T$ and the variables $\{  Z_t   \}_{t=1}^T$ satisfy the assumptions below.
Define $\text{sign}(x) = \mathbb{I}(x > 0) - \mathbb{I}(x < 0)$, where $\mathbb{I}(\cdot)$ is the indicator function.
\begin{assumption}
\label{ass:f}
In (\ref{eq:signoise}), $f_t$ is a piecewise-constant vector with an unknown number $N$ and locations $0 = \eta_0 < \eta_1 <  \ldots < \eta_N < \eta_{N+1} = T$ of change-points.
(The location $\eta_j$ is a change-point if $f_{\eta_j} \neq f_{\eta_j+1}$.)
\end{assumption}

\begin{assumption}
\label{ass:z}
\begin{enumerate}
\item[(a)] $\forall\,t$, $M(Z_t) = 0$, where $M$ is the median operator. (If the median is non-unique, we require $0 \in M(Z_t)$.)
\item[(b)] The variables $\{Z_t\}_{t=1}^T$ are sign-symmetric, i.e. $P(Z_t > 0) = P(Z_t < 0)$, $\forall\,t$.
\item[(c)] The variables $\{\mathrm{sign}(Z_t)\}_{t=1}^T$ are mutually independent.
\end{enumerate}
\end{assumption}

While Assumption \ref{ass:f} holds throughout the paper, Assumption \ref{ass:z} is not formally needed until Section \ref{sec:bounds}. Section \ref{sec:lsb} provides additional results under extra assumptions.

$\{Z_t\}_{t=1}^T$ do not have to be identically distributed, and can have arbitrary mass atoms, or none, as long as their distributions satisfy
Assumption \ref{ass:z}.
The distribution(s) of $\{  Z_t   \}_{t=1}^T$ can be unknown to the analyst, and 
we do not impose moment assumptions.
Any zero-median continuous distribution (even one with an asymmetric density function) is sign-symmetric. The requirement of the independence of $\{\text{sign}(Z_t)\}_{t=1}^T$ is weaker than that of the independence of $\{Z_t\}_{t=1}^T$ itself: e.g. if $Z_t$ is a (G)ARCH process driven by symmetric, independent innovations, then $\text{sign}(Z_t)$ is serially independent, while $Z_t$ is not.
While the results of Section \ref{sec:rnsp} require the independence of $\text{sign}(Z_t)$, those in Section \ref{sec:lsb} require the independence of $Z_t$; see Section \ref{sec:lsb} for details.

RNSP achieves the high level of generality in terms of the permitted distributions of $Z_t$ thanks to its use of the sign transformation.
The use of 0 in $\text{sign}(x)$ is critical for RNSP's objective to provide exact finite-sample coverage guarantees also for discrete distributions and continuous distributions with mass points, an aspect we discuss in Section \ref{sec:devdisc}. The sign function is a critical building block of various procedures for nonparametric change-point testing and estimation; key references include \cite{bj68}, \cite{c88} and \cite{d91}.

\subsection{(R)NSP: generic algorithm}

The generic algorithmic framework underlying both RNSP and the non-robust NSP in \cite{f21} is the same and is based on recursively searching for the shortest sub-samples in the data with globally significant deviations from the baseline model. In this section, we introduce this shared generic framework. In the following sections, we show how RNSP diverges from NSP through its use of a robust measure of deviation from the baseline model, suitable for the broad class of distributions specified in Assumption \ref{ass:z}.

We start with a pseudocode definition of the RNSP algorithm, in the form of a recursively defined function RNSP.
In its arguments, $[s,e]$ is the current interval under consideration and at the start of the procedure, we have $[s,e] = [1,T]$;
$Y$ (of length $T$) is as in the model formula (\ref{eq:signoise}); $M$ is the minimum guaranteed number of sub-intervals of $[s,e]$ drawn (unless the number of all sub-intervals of $[s,e]$ is less than $M$, in which case drawing $M$ sub-intervals would mean repetition); $\lambda_\alpha$
is the threshold corresponding to the global significance level $\alpha$ (typical values for $\alpha$ would be 0.05 or 0.1) and $\tau_L$ (respectively $\tau_R$) is a functional parameter
used to specify
the maximum extent of overlap of the left (respectively right) sub-interval of $[s,e]$ searched next after having identified a region of significance within $[s,e]$, if any.
The no-overlap case would correspond to $\tau_L = \tau_R \equiv 0$. In each recursive call on a generic interval $[s,e]$, RNSP
adds to the set $\mathcal{S}$ any globally significant local regions (intervals) of the data identified within $[s,e]$ on which $Y$ is deemed 
to depart significantly (at global level $\alpha$) from the baseline constant model.
We provide more details underneath the pseudocode below. In the remainder of the paper,
the subscript $_{[s,e]}$ relates to a constant indexed by the interval $[s,e]$ whose value will be clear from the context.

\begin{algorithmic}[1]
\Function{RNSP}{$s$, $e$, $Y$, $M$, $\lambda_\alpha$, $\tau_L$, $\tau_R$}
\If {$e-s < 1$}
\State STOP
\EndIf
\If {$M \ge \frac{1}{2}(e-s+1)(e-s)$}
\State ${M} := \frac{1}{2}(e-s+1)(e-s)$ 
\State draw all intervals $[s_m, e_m] \subseteq [s, s+1, \ldots, e]$, $m=1, \ldots, {M}$, s.t. $e_m-s_m \ge 1$
\Else
\State draw a representative (see description below) sample of intervals $[s_m, e_m] \subseteq [s, s+1, \ldots, e]$, $m=1, \ldots, {M}$, s.t. $e_m-s_m \ge 1$
\EndIf
\For{$m \gets 1,\ldots, M$}
\State $D_{[s_m, e_m]} :=$ \textsc{DeviationFromConstantModel}$(s_m, e_m, Y)$
\EndFor
\State $\mathcal{M}_0 := \arg\min_m\{ e_m - s_m\,\,:\,\, m = 1, \ldots, M;\,\,D_{[s_m, e_m]} > \lambda_\alpha    \}$
\If {$|\mathcal{M}_0| = 0$}
\State STOP
\EndIf
\State $m_0 := $\textsc{AnyOf}$(\arg\max_m\{ D_{[s_m, e_m]}\,\,:\,\, m\in \mathcal{M}_0\})$
\State $[\tilde{s}, \tilde{e}] :=$\textsc{ShortestSignificantSubinterval}$(s_{m_0}, e_{m_0}, Y, M, \lambda_\alpha)$
\State add $[\tilde{s}, \tilde{e}]$ to the set $\mathcal{S}$ of significant intervals
\State RNSP$(s, \tilde{s}+\tau_L(\tilde{s}, \tilde{e}, Y), Y, M, \lambda_\alpha, \tau_L, \tau_R)$
\State RNSP$(\tilde{e}-\tau_R(\tilde{s}, \tilde{e}, Y), e, Y, M, \lambda_\alpha, \tau_L, \tau_R)$
\EndFunction
\end{algorithmic}

The RNSP algorithm is launched by the pair of calls: $\mathcal{S} := \emptyset$; 
RNSP$(1, T, Y, M, \lambda_\alpha, \tau_L, \tau_R)$.
On completion, the output of RNSP is in the set $\mathcal{S}$; when the context requires it, we write the output as
$\mathcal{S}\{  \text{RNSP}(1, T, Y, M, \lambda_\alpha, \tau_L, \tau_R)  \}$.
We now comment on the RNSP function line by line.
In lines 2--4, execution is terminated for intervals that are too short.
In lines 5--10, a check is performed to see
if $M$ is at least as large as the number of all sub-intervals of $[s,e]$. If so, then $M$ is adjusted accordingly, and all sub-intervals are stored in $\{[s_m, e_m]\}_{m=1}^M$.
Otherwise, a sample of $M$ sub-intervals $[s_m, e_m] \subseteq [s,e]$ is drawn in which 
$s_m$ and $e_m$ are all possible pairs from an (approximately) equispaced grid on $[s,e]$ which permits at least $M$ such sub-intervals.
The ability to adjust the $M$ parameter offers the users a choice between a faster but less thorough procedure (for lower values of $M$) and a slower but more accurate one (for higher values of $M$). The reason for not necessarily using the maximum possible number of intervals is that this may increase the computation time beyond what is acceptable to the user. All examples in the paper use $M = 1000$.

In lines 11--13,
each sub-interval $[s_m, e_m]$ is checked to see to what extent the response on this sub-interval (denoted by $Y_{s_m:e_m}$) deviates from the baseline constant model.
This core step of the RNSP algorithm
will be described in more detail in Section \ref{sec:rnsp}.

In line 14, the measures of deviation obtained in line 12 are tested against threshold $\lambda_\alpha$, chosen to guarantee the global significance level $\alpha$. How to choose
$\lambda_\alpha$ is independent of the distribution of $Z_t$ if it is continuous, and there is also a simple distribution-independent choice of $\lambda_\alpha$ for discrete distributions and continuous distributions with probability masses; see Section \ref{sec:bounds}. The shortest sub-interval(s) $[s_m, e_m]$ for which the
test rejects the baseline model at global level $\alpha$ are collected in set $\mathcal{M}_0$. In lines
15--17, if $\mathcal{M}_0$ is empty, then the procedure decides that it has not found regions of significant deviations from the constant model on $[s,e]$, and stops on this interval as a consequence.
Otherwise, in line
18, the procedure continues by choosing the sub-interval, from among the shortest significant ones, on which the deviation from the baseline constant model has been the largest.
The chosen interval is denoted by $[s_{m_0}, e_{m_0}]$.

In line 19, $[s_{m_0}, e_{m_0}]$ is searched for the shortest sub-interval on which the hypothesis of the baseline model is rejected locally at a global level $\alpha$.
Such a sub-interval certainly exists, as $[s_{m_0}, e_{m_0}]$ itself has this property. The structure of this search again follows the workflow
of the RNSP procedure; it proceeds by executing lines 2--18 of RNSP, but with $s_{m_0}, e_{m_0}$ in place of $s, e$. The chosen interval is denoted by $[\tilde{s}, \tilde{e}]$.
This second-stage search is important to RNSP's pursuit to produce short intervals: indeed, if the sample of intervals $[s_m, e_m]$ contained insufficiently short intervals (perhaps because an insufficiently large $M$ was chosen), then, without the second-stage search in line 19, the intervals of significance returned by RNSP might be overly long. The second-stage search in line 19 can be seen as a guard against a small $M$, or in other words against an insufficiently fine original grid of interval endpoints.
In line 20, the selected interval $[\tilde{s}, \tilde{e}]$ is added to the output set $\mathcal{S}$.

Because of the second-stage search in line 19, pre-drawing the intervals $[s_m, e_m]$ prior to launching the RNSP procedure (rather than drawing them recursively on each current interval as it done in the algorithm)
is not an option for RNSP. Indeed, in the presence of the second-stage search, the selected interval of significance $[\tilde{s}, \tilde{e}]$ may be misaligned with the initial grid of intervals drawn, in which case
the grid of intervals to the left and to the right of $[\tilde{s}, \tilde{e}]$ must be re-drawn to avoid leaving un-examined gaps in the data.

In lines 21--22, RNSP is executed recursively to the left and to the right of the detected interval $[\tilde{s}, \tilde{e}]$. However, we optionally allow for some overlap with $[\tilde{s}, \tilde{e}]$.
The overlap, if present, is a function of $[\tilde{s}, \tilde{e}]$ and, if it involves detection of the location of a change-point within $[\tilde{s}, \tilde{e}]$, then it is also a function of $Y$.
An example of the relevance of this is given in Section \ref{sec:expost}.

\section{Robust NSP: measuring deviation from the constant model}
\label{sec:rnsp}

\subsection{Deviation measure: motivation, definition and properties}
\label{sec:principles}

The main structure of the \textsc{DeviationFromConstantModel}$(s_m, e_m, Y)$ operation is as follows: (1) 
Fit the best, in the sense described precisely later, constant model to $Y_{s_m:e_m}$. (2)
Examine the signs of the empirical residuals from this fit. If their distribution is deemed to contain a change-point (which indicates that the constant model fit is unsatisfactory on $[s_m, e_m]$ and therefore the model contains a change-point on that interval), the value returned by \textsc{DeviationFromConstantModel}$(s_m, e_m, Y)$ should be large; otherwise small.



\comment{
{\bf Desideratum B.}
The deviation measure on $[s_m, e_m]$ cannot be made smaller by proposing unrealistic constant model fits on that interval; otherwise it would be easy to force non-detection on any interval. This is an important desired property as our deviation measure will need to `try' all possible constant model fits and choose one for which the deviation measure is the smallest, to ensure that
Desideratum A (the part relating to boundedness from above by the deviation measure for the true model) is met.}

A key ingredient of our measure of deviation is
a multiresolution sup-norm introduced below, used on the signs of the input rather than in the original data domain.
Its basic building block
is a scaled partial sum statistic, defined for
an arbitrary input sequence $\{x_t\}_{t=1}^T$ by $U_{s,e}(x) = (e-s+1)^{-1/2} \sum_{t=s}^e x_t$.
We define the multiresolution sup-norm \citep{n85,l16} of an input vector $x$ (of length $T$) with respect to the interval set $\mathcal{I}$ as 
$\|x\|_{\mathcal{I}} = \max_{[s,e]\in\mathcal{I}} |U_{s,e}(x)|$.
The set $\mathcal{I}$ used in RNSP contains
intervals at a range of scales and locations.
A canonical example of a suitable interval set $\mathcal{I}$ is the set $\mathcal{I}^a$ of all subintervals of $[1,T]$. We will use $\mathcal{I}^a$ in defining the largest acceptable global probability of spurious detection. However, for computational reasons, \textsc{DeviationFromConstantModel} will use a smaller interval set (we give the details later). This will not affect the exactness of our coverage guarantees, because, naturally, if $\mathcal{J} \subseteq \mathcal{I}$, then $\|x\|_{\mathcal{J}} \le \|x\|_{\mathcal{I}}$.
We also define the restriction of $\mathcal{I}$ to an arbitrary interval $[s,e]$ as $\mathcal{I}_{[s,e]} = \{  [u,v] \subseteq [s,e]\,\,:\,\, [u,v] \in \mathcal{I}  \}$.
Note the trivial inequality
\begin{equation}
\label{eq:simpineq}
\|  x_{s:e}   \|_{\mathcal{I}^a_{[s, e]}} \le \| x   \|_{\mathcal{I}^a}
\end{equation}
for any $[s,e] \subseteq [1,T]$.
When the above multiresolution sup-norm is applied to the signs of the input, as is done in this work, rather than the original input, we refer to it as the sign-multiresolution sup-norm. When applied to the empirical residuals from a candidate constant fit on an interval, it can be viewed as a simple robust multiscale measure of data fidelity of the given candidate fit on all time scales up to the length of the interval.

We now define the deviation measure
$D_{[s_m, e_m]} := $ \textsc{DeviationFromConstantModel}$(s_m, e_m, Y)$,
\comment{returned by the \textsc{DeviationFromConstantModel} function from line 12 of the RNSP algorithm,}
which satisfies the property that if there is no change-point on the interval $[s_m, e_m]$, then it is guaranteed that
\begin{equation}
\label{eq:desid}
D_{[s_m, e_m]} \le \|  \text{sign}(Z_{s_m:e_m})    \|_{\mathcal{I}^a_{[s_m, e_m]}}.
\end{equation}
The discussion below assumes that there is no change-point in $[s_m, e_m]$. For the true constant signal $f_{s_m:e_m}$, denote $f_{[s_m,e_m]} := f_{s_m} = \ldots = f_{e_m}$.
There are only at most $2(e_m-s_m)+3$ different possible constants
$\tilde{f}_{[s_m,e_m]}$ leading to different sequences $\{   \text{sign}(Y_t - \tilde{f}_{[s_m,e_m]} )  \}_{t=s_m}^{e_m}$. To see this, sort the values of $Y_{s_m:e_m}$ in non-decreasing order to create $Y_{(1)}, Y_{(2)}, \ldots, Y_{(e_m-s_m+1)}$. Take candidate constants $\tilde{f}_{[s_m,e_m]}^{\{j\}}$, $j = 1, \ldots, 2(e_m-s_m)+3$, defined as follows.
\begin{eqnarray}
\tilde{f}_{[s_m,e_m]}^{\{1\}} & < & Y_{(1)}\quad\text{(but otherwise arbitrary)}\nonumber\\
\tilde{f}_{[s_m,e_m]}^{\{2\}} & = & Y_{(1)}\nonumber\\
\tilde{f}_{[s_m,e_m]}^{\{3\}} & = & \frac{1}{2}(Y_{(1)} + Y_{(2)})\nonumber\\
\tilde{f}_{[s_m,e_m]}^{\{4\}} & = & Y_{(2)}\nonumber\\
\tilde{f}_{[s_m,e_m]}^{\{5\}} & = & \frac{1}{2}(Y_{(2)} + Y_{(3)})\nonumber\\
& \vdots & \nonumber\\
\tilde{f}_{[s_m,e_m]}^{\{2(e_m-s_m)+2\}} & = & Y_{(e_m-s_m+1)}\nonumber\\
\label{eq:tildefj}
\tilde{f}_{[s_m,e_m]}^{\{2(e_m-s_m)+3\}} & > & Y_{(e_m-s_m+1)}\quad\text{(but otherwise arbitrary)}.
\end{eqnarray}
We have the following simple result; the proof is trivial and we omit it.
\begin{prop}
\label{prop1}
Under Assumption \ref{ass:f},
assume no change-point in $[s_m, e_m]$ and denote $f_{[s_m,e_m]} := f_{s_m} = \ldots = f_{e_m}$. Let the constants $\tilde{f}_{[s_m,e_m]}^{\{j\}}$, $j = 1, \ldots, 2(e_m-s_m)+3$ be defined as in (\ref{eq:tildefj}). There exists a $j_0 \in \{ 1, \ldots,   2(e_m-s_m)+3 \}$ such that
\begin{equation}
\label{eq:signeq}
\{   \mathrm{sign}(Y_t - f_{[s_m,e_m]} )  \}_{t=s_m}^{e_m} = \{   \mathrm{sign}(Y_t - \tilde{f}_{[s_m,e_m]}^{\{j_0\}} )  \}_{t=s_m}^{e_m}.
\end{equation}
\end{prop}

We now define our measure of deviation $D_{[s_m, e_m]}$, and prove its key property as a corollary to Proposition \ref{prop1}.
\begin{defin}
Let the constants $\tilde{f}_{[s_m,e_m]}^{\{j\}}$, $j = 1, \ldots, 2(e_m-s_m)+3$ be defined as in (\ref{eq:tildefj}). We define
\begin{equation}
\label{eq:d}
D_{[s_m, e_m]} := \min_{j\in\{1, \ldots,   2(e_m-s_m)+3\}} \|    \mathrm{sign}(Y_{s_m:e_m} - \tilde{f}^{\{j\}}_{[s_m,e_m]})      \|_{\mathcal{I}^a_{[s_m, e_m]}}.
\end{equation}
\end{defin}
$D_{[s_m,e_m]}$ tries all possible baseline constant model fits on $[s_m, e_m]$ and chooses the one that minimises the sign-multiresolution norm of the residuals, to ensure that
the finite-sample coverage guarantees hold, as we will see below.

\begin{cor}
\label{cor:d}
Under Assumption \ref{ass:f},
for any interval $[s_m, e_m]$ on which there is no change-point, we have
\begin{equation}
\label{eq:cor}
D_{[s_m, e_m]} \le \| \mathrm{sign}(Z_{s_m:e_m})    \|_{\mathcal{I}^a_{[s_m, e_m]}}.
\end{equation}
In other words, the deviation measure defined in (\ref{eq:d}) satisfies the desired property (\ref{eq:desid}).
\end{cor}
\noindent {\bf Proof.} Let the index $j_0$ be as in the statement of Proposition \ref{prop1}. We have
\begin{eqnarray*}
D_{[s_m, e_m]} & = & \min_{j\in\{1, \ldots,   2(e_m-s_m)+3\}} \|    \mathrm{sign}(Y_{s_m:e_m} - \tilde{f}^{\{j\}}_{[s_m,e_m]})      \|_{\mathcal{I}^a_{[s_m, e_m]}}\\
& \le & \|    \mathrm{sign}(Y_{s_m:e_m} - \tilde{f}^{\{j_0\}}_{[s_m,e_m]})      \|_{\mathcal{I}^a_{[s_m, e_m]}} = \|    \mathrm{sign}(Y_{s_m:e_m} - f_{[s_m,e_m]})      \|_{\mathcal{I}^a_{[s_m, e_m]}}\\
& = & \|    \mathrm{sign}(Z_{s_m:e_m})      \|_{\mathcal{I}^a_{[s_m, e_m]}}.
\end{eqnarray*}\hfill$\square$

This leads to the following guarantee for the RNSP algorithm.
\begin{theorem}
\label{th:main}
Let Assumption \ref{ass:f} hold, and 
let $\mathcal{S} = \{ S_1, \ldots, S_R     \}$ be the set of intervals returned by the RNSP algorithm. We have 
$P\left( \exists\,\,{i=1,\ldots, R}\,\,\,\forall\,\,{j=1,\ldots, N}\,\,\,   [\eta_j,\eta_j+1] \not\subseteq S_i    \right)     \le P(\| \mathrm{sign}(Z) \|_{\mathcal{I}^a} > \lambda_\alpha)$.
\end{theorem}
\noindent {\bf Proof.} On the set 
$\|\mathrm{sign}(Z) \|_{\mathcal{I}^a} \le \lambda_\alpha$, each interval $S_i$ must contain a change-point as if it did not, then by Corollary \ref{cor:d} and inequality (\ref{eq:simpineq}), 
we would have to have
\begin{equation}
\label{eq:contra}
D_{S_i} \le \|\mathrm{sign}(Z) \|_{\mathcal{I}^a} \le \lambda_\alpha.
\end{equation}
However, the fact that $S_i$ was returned by RNSP means, by line 14 of the RNSP algorithm, that $D_{S_i} > \lambda_\alpha$, 
which contradicts (\ref{eq:contra}). This completes the proof. \hfill $\square$

Theorem \ref{th:main} should be read as follows. Let $\alpha = P(\| \mathrm{sign}(Z) \|_{\mathcal{I}^a} > \lambda_\alpha)$. For a set of intervals returned by RNSP,
we are guaranteed, with probability of at least $1-\alpha$, that there is at least one change-point in each of these intervals. Therefore, $\mathcal{S} = \{ S_1, \ldots, S_R  \}$
can be interpreted as an automatically chosen set of {\em regions (intervals) of significance} in the data. In the no-change-point case ($N=0$),
the probability of obtaining one of more intervals of significance ($R\ge 1$) is bounded from above by $P(\| \mathrm{sign}(Z) \|_{\mathcal{I}^a} > \lambda_\alpha)$.
Theorem \ref{th:main} is of a finite-sample character and holds exactly and for any sample size. Moreover, it is independent of the form of the innovations $Z$. Assumptions on $Z$ will
only be needed in controlling the term $P(\| \mathrm{sign}(Z) \|_{\mathcal{I}^a} > \lambda_\alpha)$; we defer this to Section \ref{sec:bounds}.

We emphasise that Theorem \ref{th:main} does not promise to detect all the change-points, or to do so asymptotically as the sample size gets larger: this would be impossible without assumptions on the strength of the change-points (involving spacing between neighbouring change-points and the sizes of the jumps). This aspect of RNSP is investigated in Section \ref{sec:lsb}. Instead, Theorem \ref{th:main} promises that every interval of significance returned by RNSP must contain at least one change-point each, with a certain global probability adjustable by the user. Therefore, one particular implication of Theorem \ref{th:main} is that we must have
\begin{equation}
\label{eq:lowerbound}
P(N \ge R) \ge 1 - P(\| \mathrm{sign}(Z) \|_{\mathcal{I}^a} > \lambda_\alpha).
\end{equation}

The intervals of significance returned by RNSP have an ``unconditional confidence interval'' interpretation: they are not conditional on any prior detection event, but indicate regions in the data each of which must unconditionally contain at least one change in the underlying signal $f_t$, with a global probability of at least $1-\alpha$. Therefore, as in NSP \citep{f21}, RNSP can be viewed as performing ``inference without selection'' (where ``inference'' refers to producing the RNSP intervals of significance and ``selection'' to the estimation of change-point locations, absent from RNSP). This viewpoint also enables ``post-inference selection'' or ``in-inference selection'' if the exact change-point locations (if any) are to be estimated within the RNSP intervals of significance after or during the execution of RNSP.

\subsubsection{Deviation measure: discussion}
\label{sec:devdisc}


\comment{
\noindent {\bf Method of computation.}
(\ref{eq:d}) needs to be computed by manually trying out all candidate constants $\tilde{f}^{\{j\}}_{[s_m,e_m]}$.
While this may appear expensive, it is the only option, as (a) trying only `realistic' constant model fits would require an additional global parameter describing what it means for a local constant fit to be realistic, and (b) carrying out the fitting in an ad hoc way, e.g. by only fitting the empirical median of $Y_{s_m:e_m}$ (rather than each of the constants $\tilde{f}^{\{j\}}_{[s_m,e_m]}$), would violate the desired property (\ref{eq:desid}) and therefore not be able to lead to exact coverage guarantees.
}

\noindent {\bf Achieving computational savings without affecting coverage guarantees.} The operation of trying each constant $\tilde{f}^{\{j\}}_{[s_m,e_m]}$ in (\ref{eq:d}) is fast, but in order to accelerate it further, we introduce the two computational savings below, which do not increase $D_{[s_m, e_m]}$ and therefore respect the  inequality (\ref{eq:cor}) and hence also our coverage guarantees in Theorem \ref{th:main}.

\noindent {\em Reducing the set $\mathcal{I}^a_{[s_m,e_m]}$.} To accelerate the computation of (\ref{eq:d}), we replace the set $\mathcal{I}^a_{[s_m,e_m]}$ in $D_{[s_m, e_m]}$ with the set $\mathcal{I}^{lr}_{[s_m,e_m]} := \mathcal{I}^{l}_{[s_m,e_m]} \cup \mathcal{I}^{r}_{[s_m,e_m]}$
(with $l$ and $r$ standing for left and right, respectively),
where
$\mathcal{I}^{l}_{[s_m,e_m]} = \{ [s_m, s_m+1], [s_m, s_m+2], \ldots, [s_m, e_m]     \}$ and
$\mathcal{I}^r_{[s_m,e_m]} = \{ [s_m, e_m], [s_m+1, e_m], \ldots, [e_m-1, e_m]    \}$.
This reduces the cardinality of the set of intervals included in $D_{[s_m, e_m]}$ from $O((e_m-s_m)^2)$ to $O(e_m-s_m)$. As $\mathcal{I}^{lr}_{[s_m,e_m]} \subseteq \mathcal{I}^{a}_{[s_m,e_m]}$ and hence $\|\cdot\|_{\mathcal{I}^{lr}_{[s_m,e_m]}} \le \|\cdot\|_{\mathcal{I}^{a}_{[s_m,e_m]}}$, the results of Corollary \ref{cor:d} and Theorem \ref{th:main} remain unchanged for the thus-reduced $D_{[s_m,e_m]}$. On the other hand, $\mathcal{I}^{lr}_{[s_m,e_m]}$ has been defined in this particular way so as not to compromise the detection power in the piecewise-constant signal model. To see this, consider the following illustrative example. Suppose $Y_t = f_t$ (noiseless case) and $f_t = 0$ for $t = 1, \ldots, 50$ and $f_t = 1$ for $t = 51, \ldots, 100$. On $[s_m, e_m] = [1, 100]$, the baseline constant signal level fitted is $\tilde{f}_{[1,100]} = 1/2$ and we have $\mbox{sign}(Y_t - \tilde{f}_{[1,100]}) = -1$ for $t = 1, \ldots, 50$; $\mbox{sign}(Y_t - \tilde{f}_{[1,100]}) = 1$ for $t = 51, \ldots, 100$.
In this setting, the two multiresolution sup-norms:
$\|\mbox{sign}(Y_t - \tilde{f}_{[1,100]})   \|_{\mathcal{I}^{lr}_{[1,100]}}$ and $\|    \mbox{sign}(Y_t - \tilde{f}_{[1,100]})   \|_{\mathcal{I}^{a}_{[1,100]}}$
are identical, equal to $\sqrt{50}$ and achieve this value for the intervals $[1,50]$ and $[51,100]$, members of both $\mathcal{I}^{a}_{[1,100]}$ and $\mathcal{I}^{lr}_{[1,100]}$. This simple example illustrates the wider phenomenon that if there is a single change-point in $f_t$ on a generic interval $[s_m, e_m]$ under consideration, then in the noiseless case the multiresolution norm over the set $\mathcal{I}^{a}_{[s_m,e_m]}$ is maximised at one of the ``left" or ``right'' intervals in $\mathcal{I}^{lr}_{[s_m,e_m]}$, and we are happy to sacrifice potential negligible differences in the noisy case in exchange for the substantial computational savings.

\noindent {\em Limiting interval lengths.} In practice, the analyst may not be interested in excessively long RNSP intervals of significance and may therefore wish to ignore intervals for which
$e_m - s_m > L$ for a user-specified maximum length $L$.

\comment{
\noindent {\bf Unsuitability of CUSUM or similar contrasts in deviation measure.} We note that it would be impossible to replace the multiresolution sup-norm in 
$D_{[s_m, e_m]}$ by, for example, the CUSUM statistic or a similar contrast measure such as that described in \cite{elf21}. Consider, for example, the hypothetical definition of a deviation measure as follows:
$D_{[s_m,e_m]}^{inv} := \min_{j\in\{1, \ldots,   2(e_m-s_m)+3\}} |\mbox{CUSUM}\{   \mathrm{sign}(Y_{s_m:e_m} - \tilde{f}^{\{j\}}_{[s_m,e_m]})    \}|$,
where ``inv'' stands for invalid. The definition of the CUSUM statistic is well-established, see e.g. \cite{f14a} for details. This is an invalid definition as $\mathrm{sign}(Y_{s_m:e_m} - \tilde{f}^{\{1\}}_{[s_m,e_m]})$ is a vector of ones, and the CUSUM statistic returns zero for constant vectors, so $D_{[s_m,e_m]}^{inv}$ would not be able to offer detection under any circumstances as it would always equal zero. This is an example of a construction that violates Desideratum B.
}

\noindent {\bf Validity for non-continuously distributed data.} The following three aspects of RNSP ensure the validity of its coverage guarantees
in the presence of mass points in $Z_t$ or if the distribution of $Z_t$ is discrete: (a) the fact that the sign function defined in Section \ref{sec:mo}
returns zero if its argument is zero, (b) the fact that the test levels $\tilde{f}_{[s_m,e_m]}^{\{j\}}$ (definition (\ref{eq:tildefj})) are placed at data points, and (c) the
fact that these test levels are placed in between the sorted data points. Indeed, in the absence of (a), (b) or (c), it is easy to construct simple discrete distributions
of $Z_t$ for which $D_{[s_m,e_m]}$ would be spuriously large in the absence of change-points on $[s_m, e_m]$.

\subsection{Evaluation and bounds for $\|  \text{sign}(Z)   \|_{\mathcal{I}^a}$}
\label{sec:bounds}

To make Theorem \ref{th:main} operational, we need to obtain an understanding of the distribution of $\|  \text{sign}(Z)   \|_{\mathcal{I}^a}$ so we are able to choose $\lambda_\alpha$ such that
$P(\|  \text{sign}(Z)   \|_{\mathcal{I}^a} > \lambda_\alpha) = \alpha$ (or approximately so) for a desired global significance level $\alpha$.

Initially we consider $Z_t$ such that $P(\text{sign}(Z_t) = 1) = P(\text{sign}(Z_t) = -1) = 1/2$ (the general case $P(\text{sign}(Z_t) = 0) \ge 0$ is covered in the next paragraph). One simple way of determining the distribution of $\|  \text{sign}(Z)   \|_{\mathcal{I}^a}$ for any finite $T$ is by simulation; this would only need to be done once for every $T$ and the quantiles stored for fast access. Another approach is asymptotic and proceeds as follows.
From Theorem 1.1 in \cite{kw14} (which applies to sequences of serially independent symmetric Bernoulli variables as explained in Section 1.5.1 of that work; our Assumption \ref{ass:z} means that result is applicable in our context), we have
\begin{equation}
\label{eq:kab}
\lim_{T\to\infty} P(\|  \text{sign}(Z)   \|_{\mathcal{I}^a}    >     a_T + \tau / a_T) = 1 - \exp(-2 \Lambda \exp(-\tau)),
\end{equation}
where $a_T = \{   2 \log(T \log^{-1/2} T)    \}^{1/2}$ and $\Lambda$ is a constant. As the theoretical calculation of $\Lambda$ in \cite{kw14} contains an error, we use simulation over a range of values of $T$ and $\tau$ to determine a suitable value of $\Lambda$ as 0.274. The practical choice of the significance threshold $\lambda_\alpha$ then proceed as follows: (a) fix $\alpha$ to the desired level, for example 0.05 or 0.1; (b) obtain the value of $\tau$ by equating $1 - \exp(-2 \Lambda \exp(-\tau)) = \alpha$; (c) obtain $\lambda_\alpha = a_T + \tau / a_T$. While this approach is asymptotic in nature (note the limit as $T\to\infty$ in (\ref{eq:kab})), we observe that the finite-sample agreement of $P(\|  \text{sign}(Z)   \|_{\mathcal{I}^a}    >     a_T + \tau / a_T)$ with its limit in (\ref{eq:kab}) is excellent even for small sample sizes. If the user chooses to pursue this route of obtaining $\lambda_\alpha$, this will be the only asymptotic component of RNSP.

Suppose now that $P(\text{sign}(Z_t) = 0) = \rho_t \ge 0$; note that the sign-symmetry Assumption \ref{ass:z}(b) implies $P(\text{sign}(Z_t) = 1) = P(\text{sign}(Z_t) = -1) = (1 - \rho_t)/2$. Construct the variable $\tilde{Z}_t = Z_t\,\, |\,\, Z_t \neq 0$. As $P(\text{sign}(\tilde{Z}_t) = 1) = P(\text{sign}(\tilde{Z}_t) = -1) = 1/2$, the limiting statement (\ref{eq:kab}) applies to $\text{sign}(\tilde{Z}_t)$. However, we have the double inequality
\begin{equation}
\label{eq:double}
\|  \text{sign}(Z)   \|_{\mathcal{I}^a} \le \|  \text{sign}(\tilde{Z})   \|_{\mathcal{I}^a_I} \le \|  \text{sign}(\tilde{Z})   \|_{\mathcal{I}^a},
\end{equation}
with $I = [1, 2, \ldots, T_1]$, where $T_1 = |\{  t\in[1,\ldots,T]\,\, : \,\, Z_t \neq 0   \}|$. The first inequality in (\ref{eq:double}) holds because every constituent partial sum of $\|  \text{sign}(Z)   \|_{\mathcal{I}^a}$ has a corresponding larger or equal in magnitude partial sum in $\|  \text{sign}(\tilde{Z})   \|_{\mathcal{I}^a_I}$ constructed by removing the zeros from its numerator and decreasing (or not increasing) its denominator as it contains fewer (or the same number of) terms. As an illustrative example, suppose the sequence of $\text{sign}(Z_t)$ starts $-1, 0, 1, 1$. The absolute partial sum $| -1 + 0 + 1 + 1|/\sqrt{4}$, a constituent of $\|  \text{sign}(Z)   \|_{\mathcal{I}^a}$, is majorised by the absolute partial sum $| -1 + 1 + 1|/\sqrt{3}$, a constituent of $\|  \text{sign}(\tilde{Z})   \|_{\mathcal{I}^a_I}$, where the latter sum has been constructed by removing the 0 from  $-1, 0, 1, 1$ and adjusting for the number of terms (now 3 instead of 4). The second inequality in (\ref{eq:double}) holds simply because $T_1 \le T$.
The implication of (\ref{eq:double}) is that $\|  \text{sign}(Z)   \|_{\mathcal{I}^a}$ for $\rho_t \ge 0$ is majorised by $\|  \text{sign}(Z)   \|_{\mathcal{I}^a}$ for $\rho_t = 0$, the case handled by (\ref{eq:kab}). Therefore, the threshold $\lambda_\alpha$ obtained as a consequence of (\ref{eq:kab}) can also meaningfully be applied in the general case $\rho_t \ge 0$.

\section{Detection consistency and lengths of RNSP intervals}
\label{sec:lsb}

This section shows the large-sample consistency of RNSP in detecting change-points, and the rates at which the lengths of the RNSP intervals contract (relative to $T$), as $T$ increases. To simplify our technical arguments, we consider a version of the RNSP algorithm that considers all subintervals of $[1,T]$.
Our focus on i.i.d. $Z_t$'s in this section is mainly due to our ability to rely on the Dvoretzky-Kiefer-Wolfowitz inequality in the proof of Corollary \ref{cor:consist}; however, note that the finite-sample result of Theorem \ref{th:consist} holds regardless of the dependence structure of $Z_t$.
We focus on continuously-distributed $Z_t$'s as this results in notationally much less involved arguments regarding the minimum signal strength required.
We first introduce some essential notation, and then state our assumption and the result. For each change-point $\eta_j$, define
\begin{eqnarray}
\label{eq:deltaj}
\Delta_j & = & \min\{  P\{Z_t \in (-|f_{\eta_j} - f_{\eta_j+1}|/2, 0)\},   P\{Z_t \in (0, |f_{\eta_j} - f_{\eta_j+1}|/2)\},\\
\label{eq:bardj}
\bar{d}_j & = & \bar{d}_j(\lambda, \lambda_\alpha) = \left\lceil \left( \frac{2\lambda + \lambda_\alpha}{2\Delta_j}  \right)^2 +1 \right\rceil.
\end{eqnarray}
In addition, for any process $V_t$, define $\epsilon^V_t(w) = \mathbb{I}(V_t-w > 0) - P(V_t-w > 0)$.

\begin{assumption}
\label{ass:zconsist}
\begin{enumerate}
\item[(a)] The variables $\{ Z_t  \}_{t=1}^T$ are mutually independent.
\item[(b)] The variables $\{ Z_t  \}_{t=1}^T$ are identically distributed.
\item[(c)] The distribution of $Z_1$ is continuous.
\item[(d)] With the notation $\eta_0 = 0$ and $\eta_{N+1} = T$, we have 
$\eta_{j+1} - \eta_j \ge 2\bar{d}_{j+1} + 2\bar{d}_j - 2$ for $j = 1, \ldots, N-1$, and $\eta_1 - \eta_0 \ge 2 \bar{d}_1 - 1$ as well as
$\eta_{N+1} - \eta_N \ge 2\bar{d}_N - 1$.
\end{enumerate}
\end{assumption}

Our first result below is of a finite-sample nature.

\begin{theorem}
\label{th:consist}
Let Assumptions \ref{ass:f}, \ref{ass:z}(a) and \ref{ass:zconsist}(b),(c),(d) hold. On the set defined by the intersection of the events
$\|  \mathrm{sign}(Z)   \|_{\mathcal{I}^a} \le \lambda_\alpha$ and 
$\max_{s,e}  \sup_w \left|  \frac{1}{\sqrt{e-s+1}}  \sum_{t=s}^e \epsilon^Z_t(w) \right|  \le \lambda$,
a version of the RNSP algorithm that considers all intervals, executed with no overlaps and with threshold $\lambda_\alpha$, returns
exactly $N$ intervals of significance $[s_1, e_1] < \ldots < [s_N, e_N]$ such that $\eta_j \in [s_j, e_j-1]$ and $e_j - s_j +1 \le 2 \bar{d}_j$ for $j = 1, \ldots, N$.
\end{theorem}

Theorem \ref{th:consist} leads to the following corollary giving a large-sample consistency result.
\begin{cor}
\label{cor:consist}
Let the assumptions of Theorem \ref{th:consist} and Assumption \ref{ass:zconsist}(a) hold. Let $\lambda_\alpha = (1+\delta) \{ 2\log\,T   \}^{1/2}$ and $\lambda = (1+\delta) \log^{1/2}T$, for any $\delta > 0$.
Let $\mathcal{S}$ denote the set of intervals of significance $[s_1, e_1] < \ldots < [s_R, e_R]$ returned by RNSP algorithm that considers all intervals, executed with no overlaps and with threshold $\lambda_\alpha$.
We have
\[
P\left\{  R = N\quad \land\quad \forall\,\,j=1,\ldots, N\quad \eta_j \in [s_j, e_j-1]\,\,\,\land\,\,\, e_j-s_j+1 \le 2\bar{d}_j    \right\} \to 1
\] 
as $T \to \infty$.
\end{cor}

This and Corollary \ref{cor:consist2} below are the only large-sample results of the paper; the others are of a finite-sample character. Setting $\lambda_\alpha = (1+\delta) \{ 2\log\,T   \}^{1/2}$ and $\lambda = (1+\delta) \log^{1/2}T$, for any $\delta > 0$,
causes the probabilities of the events
$\|  \mathrm{sign}(Z)   \|_{\mathcal{I}^a} \le \lambda_\alpha$ and 
$\max_{s,e}  \sup_w \left|  \frac{1}{\sqrt{e-s+1}}  \sum_{t=s}^e \epsilon^Z_t(w) \right|  \le \lambda$ in Theorem \ref{th:consist} (respectively) to converge to one. Note that here, $\lambda_\alpha$ does not depend on
$\alpha$ and is of a higher order of magnitude than the ($\alpha$-dependent) $\lambda_\alpha$ of Section \ref{sec:bounds}. Paraphrasing, this is to say that we need the global significance level $\alpha$ to tend to zero
with $T$ in order to obtain large-sample consistency; a result in line with analogous results in \cite{psm17} and \cite{vbm22}.

We briefly comment on what the result of Corollary \ref{cor:consist} means for the minimum signal strength Assumption \ref{ass:zconsist}(iii), and for the localisation rates of the RNSP algorithm in detecting the change-points. If the distribution of $Z_t$ does not vary with $T$ (this section already assumes that it does not vary with $t$), and if the jump sizes $|f_{\eta_j} - f_{\eta_{j-1}}|$ are bounded from below by a positive constant independent of $j$ and $T$, then $\Delta_j$ (formula (\ref{eq:deltaj})) is also bounded from below by a positive constant independent of $j$ and $T$. By formula (\ref{eq:bardj}), the assumptions on $\lambda, \lambda_\alpha$ in Corollary \ref{cor:consist} then imply $\bar{d}_j = \Theta(\log\,T)$ (where $\Theta$ should be read ``of the exact order''). Assumption \ref{ass:zconsist}(iii) then requires that the spacings between the change-points be at least of order $\log\, T$. Corollary \ref{cor:consist} states that the length of each RNSP interval of significance, $e_j - s_j + 1$, is, with global probability approaching 1, at most of order $\log\,T$. These minimum-spacing assumptions and the implied lengths of the localisation intervals are near-optimal and the same as those in the non-robust literature, see e.g. Theorem 1 in \cite{bcf16} and Corollary 4 in \cite{f21}, as well as the associated discussions. However, the results of this section also permit $\Delta_j \to 0$ with $T$, which will, naturally, affect the above minimum-spacing requirements and localisation rates as stipulated by formulae (\ref{eq:deltaj}) and (\ref{eq:bardj}) and Assumption \ref{ass:zconsist}(iii).

The consistency of RNSP in the sense of Corollary \ref{cor:consist} implies the consistency of any pointwise estimators $\hat{\eta}_j$ contained within the RNSP intervals of significance $[s_j, e_j]$, with the localisation rate of $\hat{\eta}_j$ bounded from above by the length of the interval $[s_j, e_j]$. 
In particular, the near-optimality of the lengths of the RNSP intervals $[s_j, e_j]$ (as discussed in the preceding paragraph) automatically implies the near-optimality of the localisation rate of $\hat{\eta}_j$.
Indeed, since by Corollary \ref{cor:consist}, $[s_j, e_j-1]$ is guaranteed to contain $\eta_j$ (on an event of high probability), for any estimator $\hat{\eta}_j \in [s_j, e_j-1]$, we must have $|\hat{\eta}_j - \eta_j| \le |e_j - s_j - 1|$ (on the same event). In other words, the rate with which $[s_j, e_j]$ contract (relative to $T$) is inherited by any estimator $\hat{\eta}_j \in [s_j, e_j - 1]$; this applies even to naive estimators constructed e.g. as the middle points of their corresponding RNSP intervals $[s_j, e_j]$, i.e. $\hat{\eta}_j = \lfloor (s_j + e_j)/2  \rfloor$. More refined estimators, e.g. one based on the CUSUM maximisation of the signs of the data around their median \citep{ss75} within each RNSP interval, can also be used and will also automatically inherit the consistency and rate. Sections \ref{sec:num} and  \ref{sec:data} illustrate both of these estimators.
Trivially, in light of Corollary \ref{cor:consist}, the set of estimated change-points $\{  \hat{\eta}_j \}_{j=1}^R$ is consistent in the Hausdorff measure for $\{ {\eta}_j \}_{j=1}^N$, the set of true change-points in $f_t$.

Yet another implication of Theorem \ref{th:consist} appears below.

\begin{cor}
\label{cor:consist2}
Let the assumptions of Corollary \ref{cor:consist} hold. Let $\lambda_\alpha$ be such that 
$P(\| \mathrm{sign}(Z) \|_{\mathcal{I}^a} \le \lambda_\alpha) \ge 1 - \alpha$ and 
let $\lambda = (1+\delta) \log^{1/2}T$, for any $\delta > 0$.
Let $\mathcal{S}$ denote the set of intervals of significance $[s_1, e_1] < \ldots < [s_R, e_R]$ returned by RNSP algorithm that considers all intervals, executed with no overlaps and with threshold $\lambda_\alpha$.
We then have
\[
\lim\inf_{T\to\infty} P\left\{  R = N\quad \land\quad \forall\,\,j=1,\ldots, N\quad \eta_j \in [s_j, e_j-1]\,\,\,\land\,\,\, e_j-s_j+1 \le 2\bar{d}_j    \right\} \ge 1 - \alpha.
\] 
\end{cor}

The computation of the deviation measure $D_{[s,e]}$ for all subintervals $[s,e]$ of $[1, T]$, as the results of this section require, is a $O(T^3)$ operation.
The cubic complexity should not surprise in the context of a robust method that considers all intervals, as there are $O(T^2)$ intervals to consider and $O(T)$ binary exceedance levels within each interval. In practice, we use three devices to reduce the computational complexity of RNSP: (a) using a fixed (i.e. unchanging with $T$) value of $M$, (b) restricting the length of intervals under consideration to $\le L$, and (c) reducing the set ${\mathcal I}^a_{[s,e]}$. Items (b) and (c) are described in more detail in Section \ref{sec:devdisc}.

\section{Numerical illustrations}
\label{sec:num}




In this section, we demonstrate numerically that the guarantee offered by Theorem \ref{th:main} holds for RNSP in practice over a variety of homogeneous and heterogeneous models for which the variables $Z_t$ satisfy Assumption \ref{ass:z}. We also investigate the circumstances under which similar guarantees are not offered by H-SMUCE \citep{psm17}, MQS \citep{vbm22} or the self-normalised version of NSP (SN-NSP), suitable for heterogeneous data \citep{f21}. In this section, we use the acronyms RNSP and SN-NSP to denote the versions of these respective procedures with no interval overlaps, i.e. $\tau_L = \tau_R = 0$. Later in this section, we introduce notation for versions with overlaps. Both RNSP and SN-NSP use $M = 1000$ intervals, the default setting. For H-SMUCE, the function call we use is \verb+stepR::stepFit(x, alpha=0.1, family="hsmuce", confband=TRUE)+. The \verb+type+ parameter in MQS 
specifies the loss function for their final estimate with multiscale constraints; we denote by MQS-R the result of
\verb+mqs::mqse(x, alpha=0.1, conf=TRUE, type="runs")+ and by MQS-K the result of
\verb+mqs::mqse(x, alpha=0.1, conf=TRUE, type="koenker")+. We use the following package versions: \verb+mqs+ v1.0, \verb+stepR+ v2.1-8, \verb+nsp+ v1.0.0.

We begin with null models, by which we mean models (\ref{eq:signoise}) for which $f_t$ is constant throughout, i.e. $N = 0$. For null models, Theorem \ref{th:main} promises that RNSP at level $\alpha$ returns no intervals of significance with probability at least $1-\alpha$. In this section, we use $\alpha = 0.1$. There are analogous parameters in H-SMUCE, MQS and SN-NSP, and they are also set to 0.1. However, while in both RNSP and SN-NSP, the parameter $\alpha$ is responsible for finite-sample coverage guarantees (for both null and non-null models), in MQS and H-SMUCE it is responsible for similar but only asymptotic coverage guarantees, as $T \to \infty$. For H-SMUCE, this latter point is clarified (jointly) in Theorem 5 of \cite{psm17} and in Section A.1 of the online supplement to that work, and for MQS -- in Theorem 2.3 of \cite{vbm22} and in Section S.4 of the online supplement to that work.

\begin{table}
{\small
\centering
\begin{tabular}{ |c|l| } 
\hline
model name & sample path execution in R \\
\hline\hline
Plain Gauss & \verb+rnorm(100)+\\
Plain Gauss Long & \verb+rnorm(1000)+\\
Plain Poisson & \verb+as.numeric(rpois(200, 1))+\\
Heterogeneous Gauss & \verb+c(rep(1, 100), rep(8, 50), rep(1, 100)) * rnorm(250)+\\
Symmetric Bernoulli & \verb+as.numeric(rbinom(200, 1, .5))+\\
Plain Cauchy & \verb+rcauchy(100, 0)+\\
Mix 1 & \verb+sample(3, size=300, replace=TRUE, prob=c(.35, .3, .35)) -> xx+\\
 & \verb+  xx[xx != 2] <- rnorm(sum(xx !=2 ))+\\
Mix 2 & \verb|rpois(200, 5)+rnorm(200)/30|\\
 \hline
\end{tabular}
\caption{Null models for the comparative simulation study in Section \ref{sec:num}.\label{tab:models}}
}
\end{table}

\begin{table}
{\small
\centering
\begin{tabular}{ |c|c|c|c|c|c| } 
\hline
model & RNSP & H-SMUCE & MQS-R & MQS-K & SN-NSP \\
\hline\hline
Plain Gauss & 100.0  & 99.0 &  99.5 & 99.5 & 100.0\\
Plain Gauss Long & 100.0  & 96.5 & 98.5 & 98.5 & 100.0\\
Plain Poisson & 98.5  & 0.0 & 99.5 & 99.5 & 1.0\\
Heterogeneous Gauss & 99.5 & 98.0 & 99.5 & 99.5 & 91.0\\
Symmetric Bernoulli & 95.5 & 0.0 & 99.5 & 99.5 & 35.0\\
Plain Cauchy & 99.5 & 100.0 & 98.0 & 98.0 & 99.5\\
Mix 1 & 100.0  & 0.0 & 99.0 & 99.0 & 99.5\\
Mix 2 & 98.5 & 88.0 & 96.5 & 96.5 & 100.0\\
 \hline
\end{tabular}
\caption{Percentage of times, out of 200 simulated sample paths of each null model, that the respective method indicates no intervals of significance at level $\alpha = 0.1$ (nominal coverage = 90\%).\label{tab:nullres}}
}
\end{table}

The null models used are listed in Table \ref{tab:models}, and Table \ref{tab:nullres} shows the associated results. RNSP, MQS-R and MQS-K keep the nominal size well across all the models considered, returning no intervals of significance at least 95\% of the time in all situations.
H-SMUCE behaves correctly for the three Gaussian models, but fails for the discrete distributions and model Mix 1, which contains mass points. It is unexpectedly successful in the Plain Cauchy model, but this is perhaps because it has very limited detection power in the Cauchy model with change-points (more on this model below). It also underperforms slightly for model Mix 2, which is continuous (and within the domain of attraction of the Gaussian distribution) but multimodal.
SN-NSP fails for the discrete distributions, which is a consequence of the (asymptotically guaranteed) closeness of the self-normalised deviation measure to the appropriate functional of the Wiener process not kicking in in these instances (due to the relatively small sample sizes).

We now discuss performance for signals with change-points ($N > 0$). Table \ref{tab:models2} defines our models; the model labelled MQS.easy is taken from \url{https://github.com/ljvanegas/mqs/blob/master/mqs.ipynb} and MQS.hard and MQS.vhard are its lower signal-to-noise versions. Theorem \ref{th:main} promises that any intervals of significance returned by RNSP at levels $\alpha$ are such that, with probability at least $1-\alpha$, they each contain at least one true change-point. In addition to RNSP, H-SMUCE, MQS-R, MQS-K and SN-NSP, we also test versions of RNSP and SN-NSP with the following overlap functions:
\begin{eqnarray}
\tau_L(\tilde{s}, \tilde{e}) & = & \lfloor (\tilde{s} + \tilde{e})/2 \rfloor - \tilde{s},\nonumber\\
\label{eq:overlap}
\tau_R(\tilde{s}, \tilde{e}) & = & \lfloor (\tilde{s} + \tilde{e})/2 \rfloor +1 - \tilde{e}.
\end{eqnarray}
This setting means that upon detecting a generic interval of significance $[\tilde{s},\tilde{e}]$ within $[s,e]$, the RNSP and SN-NSP algorithms continue on the left interval
$[s, \lfloor (\tilde{s} + \tilde{e})/2 \rfloor ]$ and the right interval $[\lfloor (\tilde{s} + \tilde{e})/2 \rfloor +1, e]$ (recall that the no-overlap case results uses the left
interval $[s, \tilde{s}]$ and the right interval $[\tilde{e}, e]$). We denote the versions of the two procedures with the overlaps as above by RNSP-O and SN-NSP-O,
respectively. As before, we set $\alpha = 0.1$ for all methods tested.

For each model and method tested, we evaluate the following metrics, which collectively promote the detection of genuine, and the non-detection of spurious, intervals of significance: [coverage] the empirical coverage (i.e. whether at least $(1-\alpha)100\%$ of the simulated sample paths are such that any intervals of significance returned contain at least one true change-point each); [prop. gen. int.] if any intervals are returned, the proportion of those that are genuine (i.e. the proportion of those intervals returned that contain at least one true change-point); [no. gen. int.] the number of genuine intervals (i.e. the number of those intervals returned that contain at least one true change-point); and [av. gen. int. len.] the average length of genuine intervals (i.e. the average length of those intervals returned that contain at least one true change-point).
No single measure describes the performance of (any) method accurately, but the four measures in conjunction provide a clear picture. As a cartoon illustration of how naive solutions are able to skew some of these measures but not the others, consider a putative interval estimator that always returns the longest possible interval of $[1,T]$. For a model with 
$\ge 1$ change-points, ``coverage'' and ``prop. gen. int'' will return 100 and 1, respectively. However, this naive solution will be penalised by the next two measures, ``no. gen. int." and ``av. gen. int. len.". For models with a single change-point, the $[1, T]$ solution will return an interval that in most cases will be unreasonably long, and this will be picked up by the ``av. gen. int. len." measure.
In addition, for models with more than one change-point, ``no. gen. int." (equal to one) will be inaccurate.

For completeness, we also show the Mean-Square Errors (MSEs) between the reconstructed signal $\hat{f}_t$ and the truth $f_t$.
Given an increasingly sorted set $\{  \hat{\eta}_j \}_{j=1}^R$ of pointwise change-point
estimators (for any method), and defining in addition $\hat{\eta}_0 = 0, \hat{\eta}_{R+1} = T$, a natural estimate of the signal $f_t$ is a piecewise-constant vector $\hat{f}_t$ such that 
$\hat{f}_t = \text{med}(Y_{\hat{\eta}_j+1}, \ldots, Y_{\hat{\eta}_{j+1}})$ for $t = {\hat{\eta}_j+1}, \ldots, \hat{\eta}_{j+1}$, $j = 0, \ldots, R$, where med is the empirical median operator.
HSMUCE comes with its own pointwise change-point estimates. MQS, NSP or RNSP do not automatically provide pointwise location estimates. For these three methods, we use two different approaches to producing pointwise change-point location estimates within each interval of significance: (a) we take the mid-points of the respective intervals of significance and (b) we take the argument-maximum of the absolute value of the CUSUM statistic of the signs of the data around the empirical median within each interval of significance.
(Midpoints of RNSP intervals of significance, while seemingly appearing ad hoc as pointwise estimates of change-point location, often behave well empirically, which may be due to the fact that RNSP pursues short intervals, and those tend to be symmetric around the true change-points as this offers the same amount of evidence on either side of the change-points -- hence the frequent empirical closeness of RNSP interval midpoints to the truth.)

\begin{table}
{\scriptsize
\centering
\begin{tabular}{ |c|l| } 
\hline
model name & sample path execution in R \\
\hline\hline
Blocks & \verb+blocks <- c(rep(0, 204), rep(14.63795, 62), rep(-3.659487, 41), rep(7.318975, 164),+ \\
& \verb+  rep(-7.318975, 40), rep(10.97846, 308), rep(-4.391385, 82), rep(3.293539, 430),+ \\
& \verb+  rep(19.02933, 225), rep(7.684923, 41), rep(15.36985, 61), rep(-3.250278e-15, 390))+\\
& \verb|blocks + 10 * rnorm(2048)| \\
Cauchy & \verb+c(rcauchy(100, 1), rcauchy(100, 2), rcauchy(100, 1))+\\
Bursts & \verb+(c(rep(1, 200), rep(3, 80), rep(1, 200), rep(3, 80), rep(1, 200),+\\
& \verb+  rep(4, 40)) * rnorm(800))^2+\\
Poisson  & \verb+as.numeric(rpois(350, c(rep(1, 50), rep(4, 50), rep(10, 50), rep(2, 200)))+\\
MQS.easy & \verb+f <- c(rep(-0.25, 690), rep(0.12, 435), rep(1.07, 85), rep(-0.53, 255), rep(0.60, 75),+\\
& \verb+ rep(-0.69, 155), rep(-0.10, 790))+\\
& \verb+z <- c(rnorm(350, sd = sqrt(1)), rt(1540-350, df = 3)/sqrt(3)*sqrt(0.1),+\\
& \verb+ (rchisq(length(mu)-1540, 1)-qchisq(0.5, 1))/sqrt(2)*sqrt(0.05))+\\
& \verb|f + z|\\
MQS.hard & \verb|f + 5 * z|\\
MQS.vhard & \verb|f + 10 * z|\\
 \hline
\end{tabular}
\caption{Non-null models for the comparative simulation study in Section \ref{sec:num}.\label{tab:models2}}
}
\end{table}




\begin{table}
{\scriptsize
\centering
\begin{tabular}{ |c|c|c|c|c|c|c|c|c| } 
\hline
model & attribute & RNSP & RNSP-O & H-SMUCE & MQS-R & MQS-K & SN-NSP & SN-NSP-O \\
\hline\hline 
& coverage                    &  100 & 100 & 30 & 14 & 14 & 100 & 100\\
Blocks & prop. gen. int. &  1  & 1 &  0.83  & 0.75 & 0.75 &  1 &  1\\
(11 cpts) & no. gen. int.                &   6.17  & 8.19 &  4.98 & 4.77 & 4.77 &  4.93 &  6.10\\
& av. gen. int. len.          &  90.65 & 101.07 & 94.72 & 88.71 & 88.71 & 160.37 & 182.83\\
& MSE (midpoint)          &     15.47  & 11.87 & $12.381^{(*)}$ & 15.903 & 15.903 &  16.40 &  15.5       \\
& MSE (CUSUM)          &      11.92   & 6.94 &  - & 13.008 & 13.008 & 14.01 &  11.5      \\
 \hline\hline
& coverage                     &  100 & 100 &   100 &  71 &  70 & 100 &  99.5\\
Cauchy & prop. gen. int. &   1  &  1  &    1 &  0.61 &  0.59 &  1 &  0.99\\
(2 cpts) & no. gen. int.                 &     0.62 &   0.78  &    0.005 &  0.58 &  0.56  & 0.27 &  0.3\\
& av. gen. int. len.           &   119.16 & 124.32  &  170 & 116.59 & 116.28 & 159.58 & 159.79\\
& MSE (midpoint)          &      0.210  &  0.193 &  $0.235^{(*)}$ &  0.210 &  0.210 &  0.229 &  0.227        \\
& MSE (CUSUM)          &       0.209  &  0.194 &   -  & 0.208 &  0.208 &  0.225 &  0.224       \\
 \hline\hline
& coverage                      & 100 & 100 &  43.5 &  25.5 & 27 & 100 & 100\\
Bursts & prop. gen. int. &  1 &   1 &   0.79 & 0.6 &  0.61 &  1 &  1 \\
(5 cpts) & no. gen. int.                &   3.10 &  4.44  & 3.19 & 2.14 & 2.17 &  4.18 &  5.36\\
& av. gen. int. len.         &   100.79 & 107.64 & 92.85 & 94.42 & 94.40 & 111.61 & 108.81 \\
& MSE (midpoint)          &     20.1 &  16.73 & $15.935^{(*)}$ & 20.803 & 20.770 &  18.22 & 14.08       \\
& MSE (CUSUM)          &      17.4  & 12.37 & - & 17.722 & 17.673 & 13.51 &  9.97      \\
 \hline\hline
 & coverage &  100 & 100 &   0 & 76.5 & 76 & 0 &  0 \\
Poisson & prop. gen. int. & 1 &   1 &   0.1 & 0.87 &  0.86 & 0.07 &  0.07\\
(3 cpts) & no. gen. int. &   2.9 &   3 &   1.48 &  2.37 &  2.35 &  0.85 &  1.02\\
& av. gen. int. len. &   36.99  & 37.48 &  19.61 & 32.1 & 31.97 & 33.61 & 32.69\\
& MSE (midpoint)          &     0.716  &  0.636 &  $1.2322^{(*)}$ &  1.165 &  1.171  & 2.5122 &  2.2460       \\
& MSE (CUSUM)          &      0.339  &  0.204 &  -  & 0.842  & 0.860 & 2.1320 & 1.9296       \\
 \hline\hline
& coverage &   100  &  100 & 53 &   100 &   100 &  99.5 &   99 \\
MQS.easy & prop. gen. int. &  1  &   1 &  0.92  &   1  &   1  &  1  &   1\\
(6 cpts) & no. gen. int.  &   6  &   6 &  5.99  &    6  &  6  &  5.7  &  6.2\\
& av. gen. int. len.   &  45  &   46 &  37.33  &   37  &  37 & 125 &  122.5\\
& MSE (midpoint)          &      0.00272  &  0.00271 & $0.0011^{(*)}$ &  0.00279 &  0.00279 &  0.031 &  0.0225      \\
& MSE (CUSUM)          &       0.00076  &  0.00072 &  -  & 0.00084 &  0.00084 &  0.012  & 0.0013        \\
\hline\hline
& coverage &  100 &  100 &  33.5 &  38 &  38 &  99 &    99 \\
MQS.hard & prop. gen. int. &  1 &  1 &  0.66 &   0.73  &  0.73 &  0.99  &   1 \\
(6 cpts) &  no. gen. int. &   3.5  &   4.8  &  2.13 &   2.94  &  2.94  &  1.67  &   2.7\\
& av. gen. int. len. &   216.6 &  194.6 & 282.97 & 220.89 & 220.89 & 464.08 &  605.3\\
& MSE (midpoint)          &    0.081  &  0.059 &  $0.10^{(*)}$  &  0.080 &   0.080 &  0.11 &  0.094       \\
& MSE (CUSUM)          &     0.069  &  0.036 &   -  &  0.069  & 0.069 &  0.11 &  0.080       \\
\hline\hline
& coverage &  100 &    100 &  47.5 & 63 &  63 &  99 &   99 \\
MQS.vhard & prop. gen. int. &  1 &    1 &  0.58 &  0.63 &  0.63  &   0.99  &  0.99\\
(6 cpts) & no. gen. int. &     0.69  &    1 &   0.79 &   0.77 &   0.77 &    0.97  &   1.14\\
& av. gen. int. len.   &  422.62  &  485 & 722.08 & 501.86 & 501.86 & 1149.57 & 1179.74\\
& MSE (midpoint)          &      0.120  &  0.115 &   $0.128^{(*)}$  &  0.120 &  0.120 &   0.129  &  0.128        \\
& MSE (CUSUM)          &      0.119  &  0.113  & -  & 0.121  &  0.121  &   0.133  &   0.133     \\
\hline
 \end{tabular}
\caption{Results for each model+method combination, out of 200 simulated sample paths: ``coverage'' is the percentage of times that the respective model+method combination did not return a spurious interval of significance; ``prop. gen. int." is the average proportion of genuine intervals out of all intervals returned, if any (if none are returned, the corresponding 0/0 ratio is ignored in the average); ``no. gen. int." is the average number of genuine intervals returned; ``av. gen. int. len." is the average length of a genuine interval returned in the respective model+method combination;
``MSE ($\cdot$)'' is the MSE of $\hat{f}_t$ constructed as described in the text, with the respective pointwise change-point estimation in brackets. $^{(*)}$ note HSMUCE uses its own pointwise change-point estimation.
Significance level $\alpha = 0.1$ (nominal coverage = 90\%).
All numbers rounded to two decimal digits except when such rounding takes a positive number to zero.\label{tab:models2res}}
}
\end{table}

Table \ref{tab:models2res} shows the results. H-SMUCE does not perform well in any scenario, not even in the Blocks model, an instance of the homogeneous Gaussian model, a simple sub-class of the heterogeneous Gaussian model class for which it was specifically designed (where it achieves the coverage of 30, well short of the expected 90). Its coverage of 100 in the Cauchy model is an artefact of the fact that it does not achieve almost any detections over the 100 simulated sample paths (so there are also no spurious detections).

With the exception of MQS.easy, the least challenging model, MQS frequently produces spurious intervals in signals with change-points: the coverage figures for MQS are well short of 90\% in all but one models tested.
On the upside, in MQS.easy (the only set-up in which MQS achieves the correct coverage), the average length of genuine intervals is around 20\% below that of RNSP.

RNSP and RNSP-O significantly outperform SN-NSP and SN-NSP-O in five out of the seven scenarios tested, the only exception being Bursts and MQS.vhard.
In Blocks and Cauchy, the RNSP methods achieve more detections and shorter intervals of significance (so better localisation). In Poisson, in addition, they achieve much better coverage (the SN-NSP methods are misled by the discrete nature of this relatively low-intensity Poisson dataset, for which their required asymptotics do not kick in, which results in a very large number of spurious detections). However, the SN-NSP methods work better for the Bursts data in the sense that they lead to more detections. The underlying reason is that the signal level in this model is linearly proportional to the standard deviation of the noise, which particularly suits the self-normalised SN-NSP methods.
The RNSP methods are the clear winners for the MQS.easy and MQS.hard models. No method performs particularly well for the MQS.vhard model, but the SN-NSP methods achieve more detections than RNSP there, although at the price of the intervals being much longer.

With regards to the MSE (midpoint) and MSE (CUSUM) measures, the clear overall winner is the RNSP-O method with sign-CUSUM localisation; however, all four versions of the RNSP approach offer competitive performance.

\section{Data examples}
\label{sec:data}

\subsection{The US real interest rates}
\label{sec:expost}

\begin{figure}[h]
    \centering
    \begin{minipage}{.5\textwidth}
        \centering
        \includegraphics[width=\linewidth]{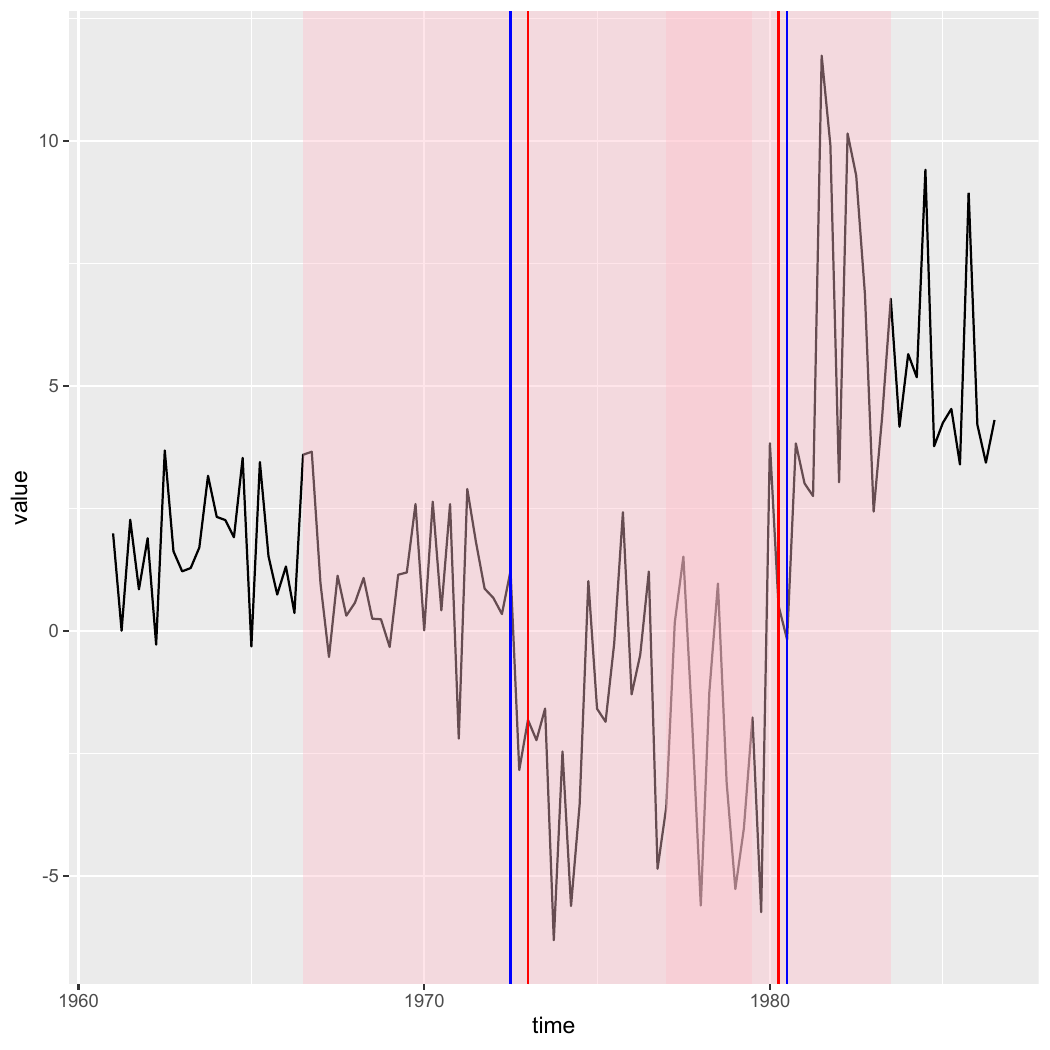}
    \end{minipage}%
    \begin{minipage}{.5\textwidth}
        \centering
        \includegraphics[width=\linewidth]{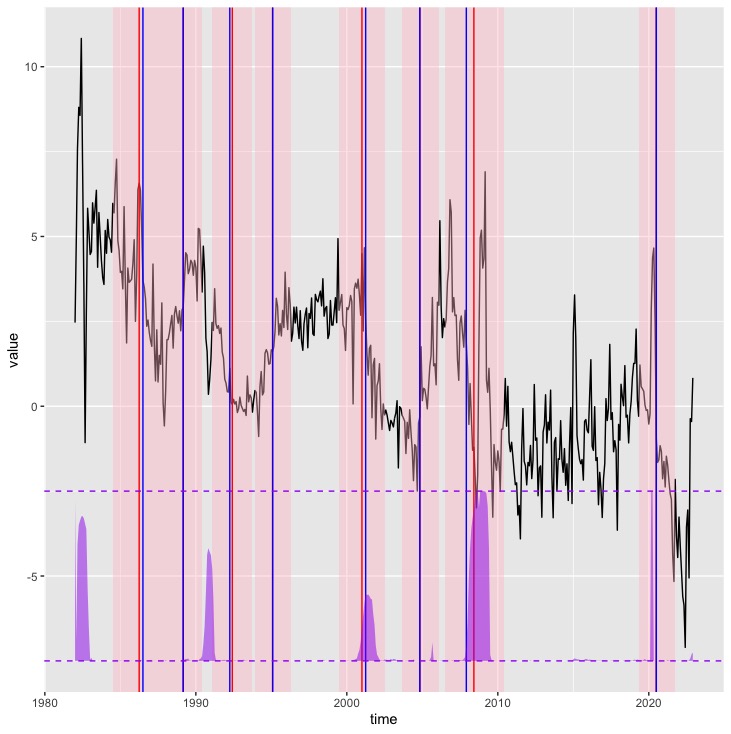}
    \end{minipage}
    \caption{Left: The US 3-month ex-post real interest rate time series (black); intervals of significance returned by RNSP (transparent pink); their midpoints (red); argument-maxima of absolute CUSUMs of signs of the data around the median in each interval of significance (blue). See Section \ref{sec:expost} for a detailed description. Right: The US 1-month real interest rate time series (black); intervals of significance returned by RNSP (transparent pink); their midpoints (red); argument-maxima of absolute CUSUMs of signs of the data around the median in each interval of significance (blue). Superimposed on the bottom of the graph is the probability of recession time series (purple), on the scale of 0 (bottom purple dashed line) to 1 (top purple dashed line). RNSP significance level $\alpha = 0.1$. See Section \ref{sec:expost} for a detailed description. \label{fig:real_dat}}
\end{figure}

\comment{

\begin{figure}[h]
\centering
  \includegraphics[width=.6\linewidth]{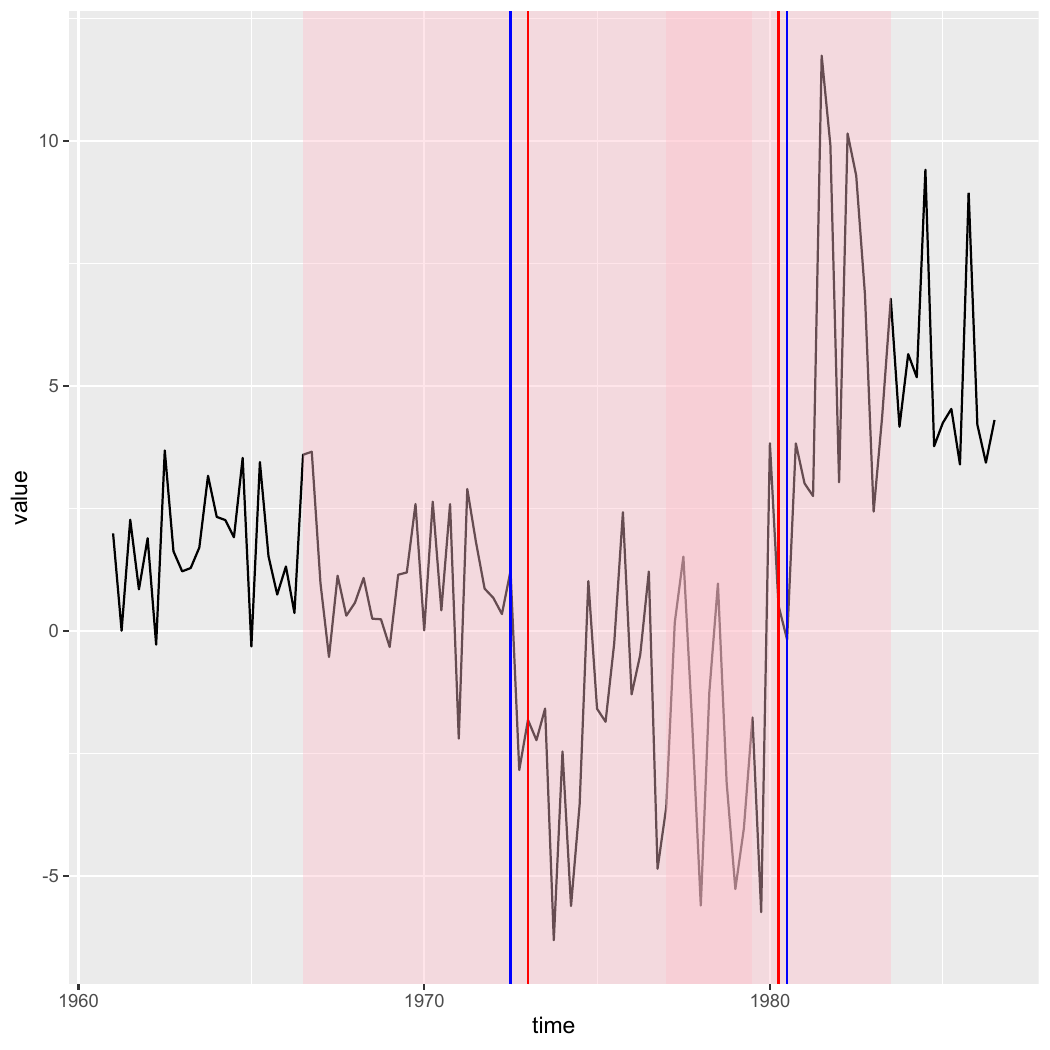}
  \caption{The US 3-month ex-post real interest rate time series (black); intervals of significance returned by RNSP (transparent pink); their midpoints (red); argument-maxima of absolute CUSUMs of signs of the data around the median in each interval of significance (blue). See Section \ref{sec:expost} for a detailed description.\label{fig:real_dat}}
\end{figure}

\begin{figure}[h]
\centering
  \includegraphics[width=.6\linewidth]{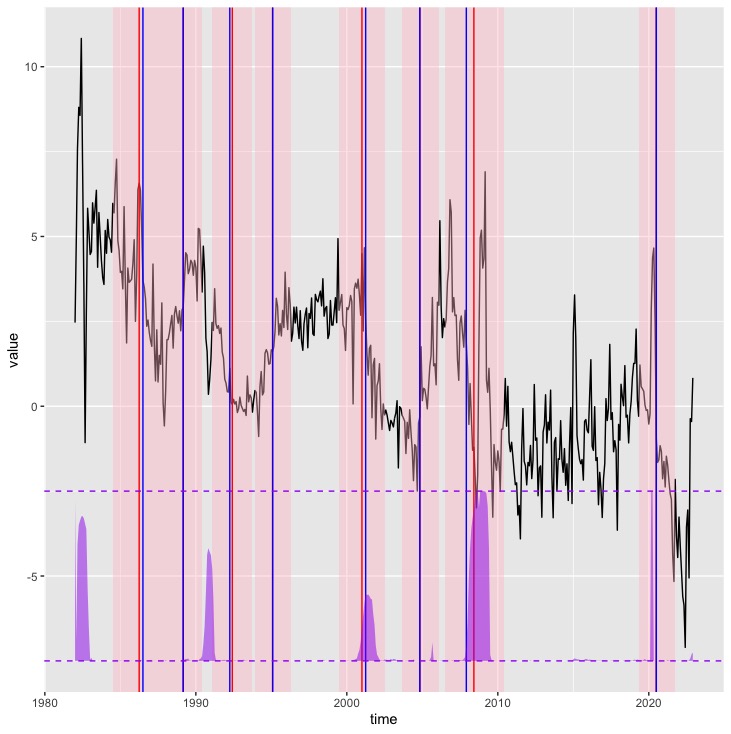}
  \caption{The US 1-month real interest rate time series (black); intervals of significance returned by RNSP (transparent pink); their midpoints (red); argument-maxima of absolute CUSUMs of signs of the data around the median in each interval of significance (blue). Superimposed on the bottom of the graph is the probability of recession time series (purple), on the scale of 0 (bottom purple dashed line) to 1 (top purple dashed line). See Section \ref{sec:expost} for a detailed description.\label{fig:real_dat1}}
\end{figure}

}

We first re-analyse the time series of US ex-post real interest rate (the three-month treasury bill rate deflated by the CPI inflation rate) considered in \cite{gp96}, \cite{bp03} and \cite{f21}.
The dataset is available at \url{http://qed.econ.queensu.ca/jae/datasets/bai001/}. The time series, shown in the left plot of Figure \ref{fig:real_dat}, is quarterly and the range is 1961:1--1986:3, so $t = 1, \ldots, T=103$.

RNSP appears to be an appropriate tool here, as the data displays heterogeneity and possibly some heavy-tailed movements towards the latter part. We run the RNSP algorithm with the default setting of $M = 1000$, with $\alpha = 0.1$ and with overlaps as defined in (\ref{eq:overlap}). The procedure returns two intervals of significance: $[23, 75]$ and $[65, 91]$. These are shown in the left plot of Figure \ref{fig:real_dat}, together with their midpoints as well as the maximisers of absolute CUSUMs of signs of the data around the median in each interval of significance. As with any RNSP execution with non-zero overlaps, one question that may be asked is whether the two intervals may be indicating the same change-point, but this, visually, is unlikely here (the reason for using non-zero overlaps is simply to provide larger samples for RNSP following the detection of the first interval; using zero overlaps means the samples are too short and RNSP with zero overlaps does not pick up the second change-point). Therefore, the solution points to a model with at least two change-points. This is consistent, or at least not inconsistent, with both \cite{gp96}, who also settle on a model with two change-points, and \cite{bp03}, who prefer a three-change-point model, not excluded by RNSP here.

The difference between those two earlier analyses and ours is that those two (a) were based on asymptotic arguments (and therefore valid asymptotically, for unspecified large samples) and (b) were conditional in the sense that the confidence regions for change-point locations in those two works were conditional on the detection event. By contrast, our analysis via RNSP has a finite-sample nature and the intervals of significance have an unconditional character. Importantly, we do not make any distributional assumptions besides independence and sign-symmetry, both of which are likely to be acceptable for this dataset. The analysis via RNSP is unaffected by the likely heterogeneity in the data.

To illustrate RNSP on a more recent dataset of a similar nature, we examine the 1-month US real interest rate, available from \url{https://fred.stlouisfed.org/series/REAINTRATREARAT1MO}. The time series, shown in the right plot of Figure \ref{fig:real_dat}, is monthly and runs from January 1982 to February 2023. We run RNSP with $M = 1000$, $\alpha = 0.1$ and no overlaps. The intervals of significance returned by RNSP appear visually plausible and it is interesting (albeit not unexpected) to observe that the periods of peaking probabilities of recession (data available from: \url{https://fred.stlouisfed.org/series/RECPROUSM156N}) from year 2000 onwards are wholly contained within RNSP intervals of significance, indicating declines in the 1-month real interest rate. The period of high probability of recession in 1982 does not appear supported in the 1-month real rate data in a way detectable to RNSP. Finally, the period of high probability of recession in 1990 appears to coincide with a falling 1-month rate but RNSP has, in this case, a visually justifiable preference for intervals just before and just after this likely recession period.

\subsection{Interest in the search term ``data science''}
\label{sec:ds}

We analyse the weekly interest in the search term ``data science'' from Google Trends, in the US state of California. The link to obtain the data
 was \url{https://trends.google.com/trends/explore?date=today%205-y&geo=US-CA&q=data%20science}.
Google Trends describe the data as follows. ``Numbers represent search interest relative to the highest point on the chart for the given region and time. A value of 100 is the peak popularity for the term. A value of 50 means that the term is half as popular. A score of 0 means there was not enough data for this term." Weeks in this data series start on Sundays and the dataset spans the weeks from w/c 28th August 2016 to w/c 15th August 2021 (so almost five years' worth of data). The observations are discrete (integers from 22 to 100), which would likely pose difficulties for the competing methods as outlined earlier.

\begin{figure}[h]
\centering
  \includegraphics[width=.6\linewidth]{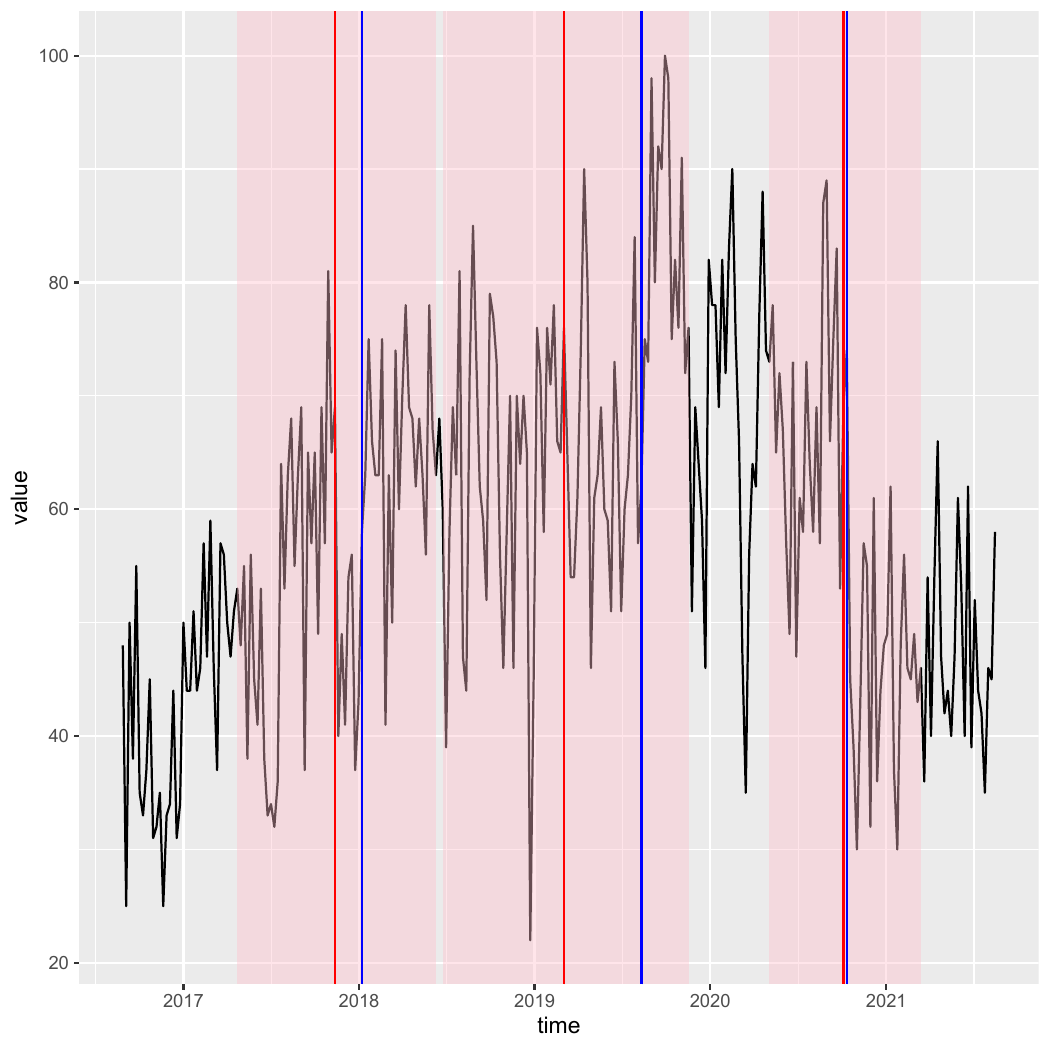}
  \caption{Weekly relative interest (top=100) in the search term ``data science'' in California in weeks from w/c 28 August 2016 to 15 August 2021 (black); intervals of significance returned by RNSP (transparent pink); their midpoints (red); argument-maxima of absolute CUSUMs of signs of the data around the median in each interval of significance (blue). RNSP significance level $\alpha = 0.1$. See Section \ref{sec:ds} for a detailed description. \label{fig:ds}}
\end{figure}

We execute the RNSP procedure with the default setting of $M = 1000$ and with $\alpha = 0.1$, with no overlaps, which returns the three intervals of significance shown in Figure \ref{fig:ds}. The intervals are: w/c 23 April 2017 -- w/c 10 June 2018 (interval 1), w/c 24 June 2018 -- w/c 17 November 2019 (interval 2), w/c 3 May 2020 -- w/c 14 March 2021 (interval 3). While intervals 1 and 2 correspond to likely increases in the median interest, interval 3 clearly corresponds to a likely decrease, and the midpoint of interval 3 (w/c 4 October 2020) visually aligns well with what seems to be a rather sudden drop of interest.

It is difficult to speculate as to possible reasons for this drop of interest. A blog post on the popular website \url{https://towardsdatascience.com} (\url{https://bit.ly/3kqwDWO})
reports that ``2020 was the first year since 2016, Data Scientist was not the number one job in America, according to Glassdoor's annual ranking. That title would belong to Front End Engineer, followed by Java Developer, followed by Data Scientist." However, visually, a similar decline in interest is observed e.g. in the analogous Google Trends series for the term ``Java" (not shown).

\comment{

Have a look at ourworldindata.org.

Also check examples from the bottom part of the \verb+NSP_robust.R+ script.

Internet traffic data? Check Mikosch for other heavy-tailed examples.

Climate data? See \url{https://arxiv.org/pdf/2106.12180.pdf}.
}

\section{Discussion}
\label{sec:disc}

{\em Other quantiles.}
Note that Corollary \ref{cor:d}, crucial to the success of RNSP, can be rewritten as 
\begin{equation}
\label{eq:cor2}
D_{[s_m, e_m]} \le \| \mathrm{sign}\{Y_{s_m:e_m} - Q_{1/2}(Y_{s_m:e_m})\}    \|_{\mathcal{I}^a_{[s_m, e_m]}},
\end{equation}
where $Q_q(\cdot)$ is the population $q$-quantile. However, the left-hand side of (\ref{eq:cor2}) was not specifically constructed with $q = 1/2$ in mind and therefore the inequality
$D_{[s_m, e_m]} \le \| \mathrm{sign}\{Y_{s_m:e_m} - Q_{q}(Y_{s_m:e_m})\}    \|_{\mathcal{I}^a_{[s_m, e_m]}}$
is true for any $q \in (0,1)$. This shows that RNSP can equally be used for significant change detection in any quantile, and not just the median. However, for $q \neq 1/2$, the challenge is to obtain the null distribution of $\| \mathrm{sign}\{Y - Q_{q}(Y)\}    \|_{\mathcal{I}^a}$. Even if this challenge is overcome (e.g. by simulation), RNSP as defined in this work may not be effective for change detection in quantiles ``far'' from the median, due to the particular way in which $D_{[s_m, e_m]}$ is constructed (involving minimisation over all levels). For RNSP to be a successful device for change detection in other quantiles, the definition of $D_{[s_m, e_m]}$ would have to be modified to only minimise over `realistic' candidate levels not far from the population $q$-quantile under the local null.

{\em Set of feasible signals.} It is convenient to define the set of feasible signals $f^0_t$ at level $\alpha$ with respect to the algorithmic execution $\text{RNSP}(1, T, \cdot, M, \lambda_\alpha, \tau_L, \tau_R)$ by
${\mathcal F}_\alpha = \{ f^0\,\, : \,\, \mathcal{S}\{ \text{RNSP}(1, T, Y - f^0, M, \lambda_\alpha, \tau_L, \tau_R)\} = \emptyset\}$,
where $\lambda_\alpha$ is such that $P(\| \mathrm{sign}(Z) \|_{\mathcal{I}^a} \le \lambda_\alpha) \ge 1 - \alpha$. That is, ${\mathcal F}_{\alpha}$ is the set of postulated signals $f^0$ such that, on fitting $f^0$ to the data and obtaining the empirical residuals, RNSP cannot distinguish (at level $\alpha$) the empirical residuals from white noise.
For any candidate fit $f^0$, it is straightforward to check whether or not it is a member of ${\mathcal F}_\alpha$ by executing $\text{RNSP}(1, T, Y - f^0, M, \lambda_\alpha, \tau_L, \tau_R)$.

\appendix

\section{Proofs}

\noindent {\bf Proof of Theorem \ref{th:consist}.}
Consider initially the case of a single change-point $\eta_1$. RNSP will, among others, consider intervals symmetric about the true change-point, i.e. $[\eta_1-d+1, \eta_1+d]$, for all appropriate $d$. Take a constant candidate fit $w$ on the interval $[\eta_1-d+1, \eta_1+d]$ and define $U_t(w) := \text{sign}(Y_t - w) = 2\mathbb{I}(Y_t - w > 0) - 1$
(the latter equality holds due to the continuity of the distribution of $Z_t$). Assume wlog $f_{\eta_1} > f_{\eta_1+1}$. We have
\begin{eqnarray}
\label{eq:devbound}
D_{[\eta_1-d+1, \eta_1+d]} & \ge & \inf_w \frac{1}{\sqrt{d}} \max \left\{  \left|\sum_{t=\eta_1-d+1}^{\eta_1} U_t(w)\right|, 
\left| \sum_{t=\eta_1+1}^{\eta_1+d} U_t(w)\right| \right\}\nonumber\\
& = & \frac{2}{\sqrt{d}} \inf_w 
\max \left\{  \left|\sum_{t=\eta_1-d+1}^{\eta_1} \mathbb{I}(Y_t - w > 0) - \frac{1}{2}\right|, 
\left| \sum_{t=\eta_1+1}^{\eta_1+d} \mathbb{I}(Y_t - w > 0) - \frac{1}{2}\right| \right\}\nonumber\\
& \ge & \frac{2}{\sqrt{d}} \inf_w 
\max \left\{  \left|\sum_{t=\eta_1-d+1}^{\eta_1} P(Y_t - w > 0) - \frac{1}{2}\right|, 
\left| \sum_{t=\eta_1+1}^{\eta_1+d} P(Y_t - w > 0) - \frac{1}{2}\right| \right\}\nonumber\\
\label{eq:devbound3}
& - & \frac{2}{\sqrt{d}} \sup_w \max \left\{ \left| \sum_{t=\eta_1-d+1}^{\eta_1} \epsilon^Y_t(w)\right|,   \left|  \sum_{t=\eta_1+1}^{\eta_1+d} \epsilon^Y_t(w) \right|   \right\}.
\end{eqnarray}
Now note
\begin{eqnarray*}
\frac{1}{\sqrt{d}} \sup_w \max \left\{ \left| \sum_{t=\eta_1-d+1}^{\eta_1} \epsilon^Y_t(w)\right|,   \left|  \sum_{t=\eta_1+1}^{\eta_1+d} \epsilon^Y_t(w) \right|   \right\} & \le &
\max_{s,e}  \sup_w \left|  \frac{1}{\sqrt{e-s+1}}  \sum_{t=s}^e \epsilon^Y_t(w) \right|\\
& = & \max_{s,e}  \sup_w \left|  \frac{1}{\sqrt{e-s+1}}  \sum_{t=s}^e \epsilon^Z_t(w) \right| \le \lambda.
\end{eqnarray*}
Continuing from
(\ref{eq:devbound3}), this implies
\begin{eqnarray}
\lefteqn{D_{[\eta_1-d+1, \eta_1+d]}}\nonumber\\
& & \ge \frac{2}{\sqrt{d}} \inf_w 
\max \left\{  \left|\sum_{t=\eta_1-d+1}^{\eta_1} P(Y_t - w > 0) - \frac{1}{2}\right|, 
\left| \sum_{t=\eta_1+1}^{\eta_1+d} P(Y_t - w > 0) - \frac{1}{2}\right| \right\} - 2 \lambda\nonumber\\
\label{eq:devbound4}
& & = \frac{2}{\sqrt{d}} \inf_w 
\max \left\{  \left|\sum_{t=\eta_1-d+1}^{\eta_1} P(Z_t > w - f_{\eta_1}) - \frac{1}{2}\right|, 
\left| \sum_{t=\eta_1+1}^{\eta_1+d} P(Z_t > w - f_{\eta_1+1}) - \frac{1}{2}\right| \right\} - 2 \lambda.\nonumber\\
\end{eqnarray}
The infimum over $w$ will be achieved if both elements of the maximum are the same (or otherwise it would be possible to alter $w$ slightly to decrease the larger of the two moduli). But this is only possible if $w \in (f_{\eta_1+1}, f_{\eta_1})$ and hence, bearing in mind that $Z_t$ is median-zero, we have
\begin{eqnarray*}
\left|\sum_{t=\eta_1-d+1}^{\eta_1} P(Z_t > w - f_{\eta_1}) - \frac{1}{2}\right| & = & d P(Z_t > w - f_{\eta_1}) - d/2 = d P\{Z_t \in (w-f_{\eta_1}, 0)\},\\
\left| \sum_{t=\eta_1+1}^{\eta_1+d} P(Z_t > w - f_{\eta_1+1}) - \frac{1}{2}\right|  & = & d/2 - d P(Z_t > w - f_{\eta_1+1}) = dP\{Z_t \in (0, w - f_{\eta_1+1})\}.
\end{eqnarray*}
Let $w_0$ be such that 
$P\{Z_t \in (w_0-f_{\eta_1}, 0)\} = P\{Z_t \in (0, w_0 - f_{\eta_1+1})\}$.
Since $(w_0 - f_{\eta_1+1}) - (w_0 - f_{\eta_1})$ = $f_{\eta_1} - f_{\eta_1+1}$, then either $f_{\eta_1} - w_0 \ge (f_{\eta_1} - f_{\eta_1+1})/2$ or $w_0 - f_{\eta_1+1} \ge (f_{\eta_1} - f_{\eta_1+1})/2$. Therefore,
\begin{eqnarray*}
P\{Z_t \in (w_0-f_{\eta_1}, 0)\} & = & P\{Z_t \in (0, w_0 - f_{\eta_1+1})\}\\
&  \ge & \min\{  P\{Z_t \in (-|f_{\eta_1} - f_{\eta_1+1}|/2, 0)\},   P\{Z_t \in (0, |f_{\eta_1} - f_{\eta_1+1}|/2)\} \} = \Delta_1.
\end{eqnarray*}
Continuing from (\ref{eq:devbound4}), we therefore have
\begin{equation}
\label{eq:suffd}
D_{[\eta_1-d+1, \eta_1+d]} \ge 2\sqrt{d} \Delta_1 - 2 \lambda.
\end{equation}
But from the definition of the RNSP algorithm (line 14), detection on an interval $[s,e]$ will occur if $D_{[s,e]} > \lambda_\alpha$. Therefore, from (\ref{eq:suffd}), detection
on $[\eta_1-d+1, \eta_1+d]$ will occur if
\begin{equation}
\label{eq:dlb}
d > \left( \frac{2\lambda + \lambda_\alpha}{2\Delta_1}  \right)^2.
\end{equation}
As RNSP looks for the shortest intervals of significance, the length of the RNSP interval will not exceed that of $[\eta_1-d+1, \eta_1+d]$, which is $2d$. From (\ref{eq:dlb}), its length will therefore be bounded from above by
$2 \left\lceil \left( \frac{2\lambda + \lambda_\alpha}{2\Delta_1}  \right)^2 +1 \right\rceil = 2\bar{d}_1$.
We now turn our attention to the multiple change-point case. 
Note that even though the RNSP interval of significance around $\eta_j$ is guaranteed to be of length at most $2\bar{d}_j$, it will not necessarily be a sub-interval of 
$[\eta_j-\bar{d}_j+1, \eta_j+\bar{d}_j]$. Therefore in order that an interval detection around $\eta_j$ does not interfere with detections around $\eta_{j-1}$ or $\eta_{j+1}$, the distances $\eta_j - \eta_{j-1}$ and $\eta_{j+1} - \eta_j$ must be suitably long, but this is guaranteed by 
Assumption \ref{ass:zconsist}(iii). This completes the proof.\hfill$\square$

\noindent{\bf Proof of Corollary \ref{cor:consist}.} The fact that $P(\|  \mathrm{sign}(Z)   \|_{\mathcal{I}^a} >  \lambda_\alpha) \to 0$
as $T \to \infty$ is a simple consequence of Corollary 1 in \cite{s95}.
We next assess and bound the magnitude of $\sup_w \left| \frac{1}{\sqrt{d}}  \sum_{t=s}^{s+d-1} \epsilon^Z_t(w)   \right|$. The Dvoretzky-Kiefer-Wolfowitz inequality (with Massart's optimal constant, see \cite{m90}) implies
\begin{equation*}
P\left(  \sup_w \left| \frac{1}{\sqrt{d}}  \sum_{t=s}^{s+d-1} \epsilon^Z_t(w)   \right|  > \lambda\right) =
P\left(  \sup_w \left| \frac{1}{d}  \sum_{t=s}^{s+d-1} \epsilon^Z_t(w)   \right|  > \lambda d^{-1/2}\right)
\le 2\exp(-2 \lambda^2).
\end{equation*}
This leads to a uniform bound via Bonferroni's correction.
\begin{eqnarray*}
P\left(  \max_{s,e}  \sup_w \left|  \frac{1}{\sqrt{e-s+1}}  \sum_{t=s}^e \epsilon^Z_t(w) \right|  > \lambda  \right) & \le & \sum_{s\le e}
P\left( \sup_w \left|  \frac{1}{\sqrt{e-s+1}}  \sum_{t=s}^e \epsilon^Z_t(w) \right|  > \lambda  \right)\\
& \le & T(T+1) \exp(-2\lambda^2).
\end{eqnarray*}
For $\lambda = (1+\delta) \log^{1/2}T$, the above tends to zero if $\delta > 0$.
This completes the proof.\hfill$\square$

\noindent{\bf Proof of Corollary \ref{cor:consist2}.}  As in the proof of Corollary \ref{th:consist}, setting $\alpha$ at this level means that
\[
P\left(\max_{s,e}  \sup_w \left|  \frac{1}{\sqrt{e-s+1}}  \sum_{t=s}^e \epsilon^Z_t(w) \right|  \le \lambda\right) \to 1
\]
as $T \to \infty$, and therefore
\[
\lim\inf_{T\to\infty}P\left(\| \mathrm{sign}(Z) \|_{\mathcal{I}^a} \le \lambda_\alpha\quad\land\quad \max_{s,e}  \sup_w \left|  \frac{1}{\sqrt{e-s+1}}  \sum_{t=s}^e \epsilon^Z_t(w) \right|  \le \lambda\right) \ge 1 - \alpha.
\]
Theorem \ref{th:consist} implies the result.\hfill$\square$

\section*{Funding acknowledgement and disclosure statement}

The author acknowledges EPSRC grant EP/V053639/1. There are no competing interests to declare.


\bibliographystyle{plainnat}

\end{document}